\documentclass[12pt,preprint]{aastex}
\usepackage{color}

\slugcomment{draft version \today, }

\shorttitle{New equations of state based on the LDM of heavy nuclei and quantum approach to light nuclei for core collapse supernova simulations}
\shortauthors{Furusawa et al.}

\begin{document}

\title{New equations of state based on the liquid drop model of heavy nuclei and quantum approach to light nuclei for core-collapse supernova simulations}
%Resolving the $z=4.4$ Quasar Host Galaxy of BRI\,1335-0417: \\
%Intimate Interaction or Massive Molecular Disk?}

\author{Shun Furusawa\altaffilmark{1}, Kohsuke Sumiyoshi\altaffilmark{2}, Shoichi Yamada\altaffilmark{1,3} and Hideyuki Suzuki\altaffilmark{4} }

\altaffiltext{1}{Advanced Research Institute for Science and Engineering, Waseda University, 3-4-1
Okubo, Shinjuku, Tokyo 169-8555, Japan}
\altaffiltext{2}{Numazu College of Technology, Ooka 3600, Numazu, Shizuoka 410-8501, Japan}
\altaffiltext{3}{Department of Science and Engineering,
 Waseda University, 3-4-1 Okubo, Shinjuku, Tokyo 169-8555, Japan}
\altaffiltext{4}{Faculty of Science and Technology, Tokyo University of Science, Yamazaki 2641, Noda, Chiba 278-8510, Japan}
\email{furusawa@heap.phys.waseda.ac.jp}

\begin{abstract}
We construct new equations of state for baryons at sub-nuclear densities for the use in core-collapse simulations of massive stars. 
The abundance of various nuclei is obtained together with thermodynamic quantities. 
A model free energy is constructed, based on the relativistic mean field theory for
 nucleons and the mass formula for nuclei with the proton number up to $\sim 1000$. 
The formulation is an extension of the previous model,
 in which we adopted the liquid drop model to all nuclei under the nuclear statistical equilibrium.
We reformulate the new liquid drop model
 so that the temperature dependences of bulk energies could be taken into account.
Furthermore, we extend the region in the nuclear chart, in which shell effects are included, 
by using theoretical mass data in addition to experimental ones.
We also adopt a quantum theoretical mass evaluation of light nuclei, which incorporates the Pauli- and self-energy shifts 
that are not included in the ordinary liquid drop model.
The pasta phases for heavy nuclei are taken into account in the same way as in the previous model.
We find that the abundances of heavy nuclei are modified by  the shell effects of nuclei and temperature dependence of bulk energies.
These changes  may have an important effect on the rates of electron captures and coherent neutrino scatterings on nuclei in supernova cores.
The abundances of light nuclei are also modified by the new mass evaluation,
 which may affect the heating and cooling rates of supernova cores and shocked envelopes. 
\end{abstract}
\section{Introduction}
%1-1EOS  role in CC
Core collapse supernovae occur at the end of the evolution of massive stars.
The mechanism of this event is not clearly understood yet
because of their intricacies (see e.g. \citet{Janka2007,Kotake2011}).
One of the underlying problems is the equations of state (EOS's) of hot and dense matter
 both at sub- and supra-nuclear densities. 
EOS provides information on compositions of nuclear matter in addition to thermodynamical quantities such as pressure, entropy and sound velocities.
The compositions  play important roles at both pre- and post- bounce phases.  
In collapsing cores, they have an influence on the rate of electron captures and neutrino coherent scatterings on nuclei,
both of which determine the evolution of the lepton fraction,
 one of the most critical ingredients for the core dynamics. 
After bounce they  affect the rates of heating and cooling through the neutrino emission and  absorption on nucleons and nuclei.

%1-2wide range need and matter phys
The EOS for the simulations of core collapse supernovae must cover a wide range of density ($10^5 \lesssim \rho_B \lesssim 10^{15} \rm{g/cm^3}$) and 
temperature ($0.1 \lesssim T \lesssim 10^{2}$~MeV), including both neutron-rich and proton-rich regimes. 
One of the difficulties in constructing the EOS is originated from the fact that depending on the density, temperature and proton fraction,
 the matter consists of either dilute free nucleons or
 a mixture of nuclei and free nucleons or strongly interacting dense nucleons. 
Another complication is the existence of the so-called nuclear pasta phases, in which nuclear shapes change from droplet to rod, slab, anti-rod and 
bubble (anti-droplet) as the density increases toward the nuclear saturation density, at which uniform nuclear matter is realized \citep{Ravenhall1983, Hashimoto1984,Oyamatsu1993,watanabe2005,Nakazato2009,Okamoto2012}. 
At high temperatures ($T \gtrsim 0.4 $~MeV), chemical equilibrium is achieved for all strong and electromagnetic reactions, 
which is referred to as nuclear statistical equilibrium, or NSE, and the nuclear composition is determined as a function of density,
 temperature, and proton fraction \citep{Timmes1999, Blinnikov2011}.
At lower temperatures, the matter composition is an outcome of preceding nuclear burnings and cannot be obtained by statistical mechanics.
In this paper we are concerned with the high temperature regime, in which the nuclear composition can be treated as a part of EOS.

At present, there are only two EOS's in wide use for the simulations of core-collapse  supernovae.
Lattimer-Swesty's EOS is based on Skyrme-type nuclear interactions and the so called compressible liquid drop model for nuclei surrounded by dripped nucleons \citep{Lattimer1991}.
The EOS by Shen et al. employs a relativistic mean field theory (RMF) to describe nuclear matter 
and the Thomas-Fermi approximation for finite nuclei with dripped nucleons.\citep{Shen1998,Hshen2011}.
It should be emphasized here that both EOS's adopt the so-called single nucleus approximation (SNA), in which only a single representative nucleus is included.
In other words, the ensemble of nuclei is ignored. \citet{Burrows1984} demonstrated that SNA is not a bad approximation for thermodynamical quantities 
such as pressure. It is not the case, however, for the weak interaction rates, since the electron capture rates are sensitive to
nuclear shell structures and the greatest contributor is not the most abundant nuclei that the single representative nuclei in SNA are
supposed to approximate \citep{Langanke2003,Hix2003}. 
In addition to the approximative calculation of heavy nuclei, only alpha particles are included in both EOS's  as a representative light nucleus.
It is predicted that not only alpha particles but deuterons, tritons and helions
are also abundant in the cooling and heating regions of cores and envelopes after bounce 
\citep{Sumiyoshi2008,Arcones2008,Hempel2012}.

In this decade, some EOS's including multi-nuclei have been formulated by different research groups.
Although all models assumed NSE, models for nuclei are different.  
Botvina's EOS \citep{Botvina04, Botvina10,Buyukcizmeci13} is a generalization of the statistical model,
which is one of the most successful models used
 for the theoretical description of multifragmentation reactions induced by heavy-ion collisions \citep{Bondorf1995}.
The calculation of the nuclear energies in this model
 is based on the liquid drop model for the mass number up to $1000$.   
However they ignored the shell effects of nuclei,
 which are important for reproducing the abundance of nuclei at low temperatures.
% %hempel
\citet{Hempel2010} utilized two mass tables, which are based on experimental data \citep{Audi2003} 
and theoretical estimation for isolated nuclei \citep{Geng2005}.
Due to the limitation of the mass tables, heavy nuclei with proton number $Z \gtrsim 100$ are not included in their NSE calculations.
They also ignored the high-density and -temperature effects on nuclear bulk and surface energies, which are explained in the  later section.
%%G shen
\citet{Shen2011} employed two different theories, the Virial expansion at low densities and SNA with the Hartree approximation at high densities.
The multi-nuclei description is employed only in the low density regime and
 some quantities such as the mass fraction of free proton are discontinuous at the transition between the two descriptions.  
%Typel
\citet{Typel2005} made an equation of state, focusing on  light nuclei.
They employed a generalized density-dependent RMF, which is applied not only to protons and neutrons but also to deuterons, tritons, helions $(=^{ 3}$He$)$ and alpha particles.  

We constructed an EOS \citep{Furusawa2011} based on the NSE description with the mass formula for nuclei up to the atomic number of 1000 
under the influence of surrounding nucleons and electrons. 
The mass formula is derived from the experimental data of nuclear binding energies and enables us to take into account nuclear
shell effects. The liquid drop model is extended to describe medium effects 
and, in particular, the formation of the pasta phases. 
%Because of this combination, 
The free energy thus obtained of the multi-component system can reproduce the 
ordinary NSE results at low densities and make a continuous transition to 
the EOS for supra-nuclear densities. 
The details of the model and comparisons with H. Shen's EOS  and Hempel's EOS are given in \citet{Furusawa2011}. 

The purpose of this study is to improve the  previous model incorporating some missing important effects
 and construct a more realistic EOS for the core-collapse supernova simulations.
As a matter of fact, our previous EOS shows unphysical jumps in the isotope distributions between the nuclei
 with the experimental mass data and those without them.
 This is demonstrated in the paper, in which we compare three different EOS's with multi-nuclei handling \citep{Buyukcizmeci}.
The main cause for this unphysical behavior is the lack of the temperature dependence in the bulk energies for the nuclei with mass data.
We hence modified the expression of bulk energies so that the temperature dependence could be incorporated in this work. %rp2
Furthermore the shell effects are taken into account only for a limited number of nuclei in our previous EOS,
 since we have used only experimental mass data \citep{Audi2003} to obtain the shell effects.
In this work, on the other hand, we utilize the theoretical mass data \citep{Koura2005} which covers 15134 nuclei that have no the experimental mass data.
In our previous EOS, we adopt the liquid drop model even for light nuclei such as deuterons, tritons, helions and alpha particles.
It is known that the liquid drop mass formula poorly reproduces the experimental mass data of the light nuclei with the mass numbers about $10$ or smaller \citep{Ghahramany2011}.
In this article, we treat light nuclei as quasi-particles immersed in dense and hot nucleons following  \citet{Typel2010}. 
%Other improvements in this article are saturation densities of individual nuclei and the calculation of the excited function of nuclei.
Other improvements in this article are saturation densities of individual nuclei and the contributions of excited states to partition functions. %rp3
In the following, we report on these new ingredients and discuss the differences from the previous version.

This article is organized as follows.
In section 2 we  overview the model free energy to be minimized and the details of new developments from the previous EOS. 
Note that the basic formulation of the model free energy and its minimization are unchanged from the previous version.
The results  are shown in section 3, with an emphasis on the differences from the previous EOS. 
The paper is wrapped up with a summary and some discussions in section~4.

\section{Formulation of the new models}
To obtain the multi-component EOS's, we construct a model free energy and minimize it with respect to the parameters included.
The matter in the supernova core at sub-nuclear densities consists of nucleons and nuclei together with electrons and photons. 
The latter two are not treated in this paper although the inclusion of them 
as ideal Fermi and Bose gases respectively is quite simple and now a routine. 
Note that the coulomb energies between protons, both inside and outside nuclei, 
and electrons are contained in the EOS
and we assume the electrons are uniformly distributed.
Neutrinos are not always in thermal or chemical equilibrium with the matter and cannot be included in  the free energies of nuclei.
Their non-equilibrium distributions should be computed with the transport equations. 

The free energy is constructed as a sum of the contributions  from free nucleons not bound in nuclei,
 light nuclei defined here as those nuclei with the proton number $Z\leq 5$, 
and the rest of heavy nuclei  with the proton and neutron numbers, $Z\leq1000$ and $N\leq1000$.
This classification of heavy or light nuclei is based on whether LDM is a good approximation in reproducing the experimental mass data or not.
It is known that the difference between LDM  and experimental masses is large for the nuclei with the mass number $A\lesssim10$ \citep{Ghahramany2011}.
We hence set the light nuclei as those with $Z\leq 5$.

 We assume that the free nucleons outside nuclei interact with themselves only in the volume that is  not occupied by other nuclei;
 light nuclei are the quasi particles whose masses are modified by the surrounding free nucleons;
 heavy nuclei are also affected by the free nucleons and electrons, depending on the temperature and density,
and contact with each other at some density and merge into pastas near the saturation densities. %rp4
The free energy of free nucleons is calculated by the RMF theory with the excluded volume effect being taken into account.
The model free energy of heavy nuclei is based on the liquid drop mass formula.
The free energy of light nuclei is approximately calculated by quantum many body theory.
In constructing the mass formula of heavy nuclei, the following issues are appropriately taken into account:
the nuclear masses at low densities and temperatures should be equal to those of isolated nuclei  in vacuum 
and the shell energies of nuclei are crucially important to reproduce the ordinary NSE (e.g. \citet{Timmes1999});
one should take into account the effect that the nuclear bulk, shell, Coulomb and surface energies
 are affected by the free nucleons and electrons at high densities and temperatures;
furthermore the pasta phases near the saturation densities should be also accounted for to ensure a continuous transition to uniform matter.
Only the bubble phase is explicitly considered in the Coulomb and surface energies and other pasta phases are just 
interpolated between the normal droplet and bubble phases.

In the following subsections, we explain the details of the free energy density expressed as
\begin{eqnarray}
\ f = f_{p,n}+\sum_{j}{n_j F_j } + \sum_i {n_i F_i}, \\
\ F_{j/i}=E^t_{j/i} + M_{j/i},
\end{eqnarray}
where $f_{p,n}$ is the free energy densities of free nucleons, $n_{j/i}$ and $F_{j/i}$ 
are the number density and free energy of individual nucleus,
 index $j$ specifying  a light nucleus with the proton number $Z_j\leq 5$ and
 index $i$ meaning a heavy nucleus with the proton number $6 \leq Z_i \leq 1000$, respectively.
$E^t_{i/j}$ and $M_{i/j}$ are the translational energies and rest masses of heavy and light nuclei.
We begin with the mass evaluation of heavy nuclei $M_i$ focusing on the modifications from our previous EOS in section~\ref{sechn}.
Then we describe the mass estimation of the light nuclei $M_j$ in section \ref{secln}.
The translational energies of  heavy and light nuclei $E^t_{j/i}$ are explained  in section \ref{sectr}.
We finally mention the evaluation of thermodynamical quantities from the free energy in section~\ref{secth}. 
Since the free energy density of free nucleons based on the RMF theory $f_{p,n}$ and
 the minimization of the total free energy densities are just the same as in the previous paper \citep{Furusawa2011},
 we briefly describe them below. 

The free energy density of free nucleons is calculated by the RMF theory with the TM1 parameter set, 
which is the same as that adopted in \citet{Shen1998}.
We take into account the excluded-volume effect:
 free nucleons can not move in the volume occupied by other nuclei, $V_N$.
Then the local number densities of free protons and neutrons are defined as $n'_{p/n}=(N_{p/n})/(V-V_N)$
 with the total volume, $V$, and the numbers of free protons, $N_p$, and  free neutrons, $N_n$.
%with the volume fraction, $\eta=(V-V_N)/V$, and the number density of free protons and free neutrons, $n_{p/n}=N_{p/n}/V$. 
Then the free energy densities  of free nucleons are defined as $f_{p,n} =(V-V_N)/V \times  f^{RMF}(n'_p,n'_n,T)$,
where $f^{RMF}(n'_p,n'_n,T)$ is the free energy density in the unoccupied volume for nucleons, $V-V_N$,
 obtained from the RMF theory at $n'_p$, $n'_n$  and temperature $T$.
 
The abundances of nuclei as a function of $\rho_B$, $T$ and $Y_p$ are obtained by minimizing the model free energy with 
respect to the number densities of nuclei and nucleons under the constraints,
\begin{eqnarray}
 n_p+n_n+\sum_j{A_j n_j} +\sum_i{A_i n_i} & = & n_B=\rho_B/m_B, \nonumber \\
 n_p+\sum_j{Z_j n_j}+\sum_i{Z_i n_i} & = & n_e=Y_p n_B,
\label{eq:cons}
\end{eqnarray}
where $n_B$ and $n_e$ are number densities  of baryon and electrons and $A_{j/i}$ and $Z_{j/i}$ are the mass and proton numbers of nucleus $j/i$.
The minimization of our free energy density is not the same as that in the ordinary NSE.
In the latter, one has only to solve the constraints, Eq.~(\ref{eq:cons}), at a given $\rho$, $T$ and $Y_p$ for two variable,
i.e., the chemical potentials of nucleons $\mu_p$ and $\mu_n$, through Saha equations.
In our case, the free energy density of nuclei depends on the local number densities of proton and neutron $n'_{p/n}$ as we will describe later. 
Thus the number densities of nuclei are not determined by $\mu_p$ and $\mu_n$ alone but they also depend on $n'_p$ and $n'_n$.
We hence have to solve  the equations relating $\mu_{p/n}$ and $n'_{p/n}$
 as well as the two constraint equations, Eq.~(\ref{eq:cons}),
to determine the four variables: $\mu_p$, $\mu_n$, $n'_p$ and $n'_n$.

%\subsubsection{free energy of nucleons}
\subsection{Mass evaluation of heavy nuclei ($Z\geq 6$)\label{sechn}} %18
The nuclear mass is assumed to be equal to the sum of shell, bulk, Coulomb and surface energies: $M_i=E_i^{Sh}+E_i^B+E_i^C+E_i^{Su}$.
In this study, we treat the shell energies separately from the bulk energies for the nuclei with mass data
unlike in the previous model, in which the shell effect was included in the bulk energies.
This is because we take into account the temperature dependence of the bulk energies for the nuclei with mass data.
Furthermore, we incorporate the dependence of the saturation density $n_{si}$ of each nucleus $i$ on $T$ and $\rho_B$. 
The formulation of Coulomb and surface energies is just identical  to the previous one.

\begin{figure}
   \begin{center}
    \begin{tabular}{c}
         \includegraphics[width=90mm]{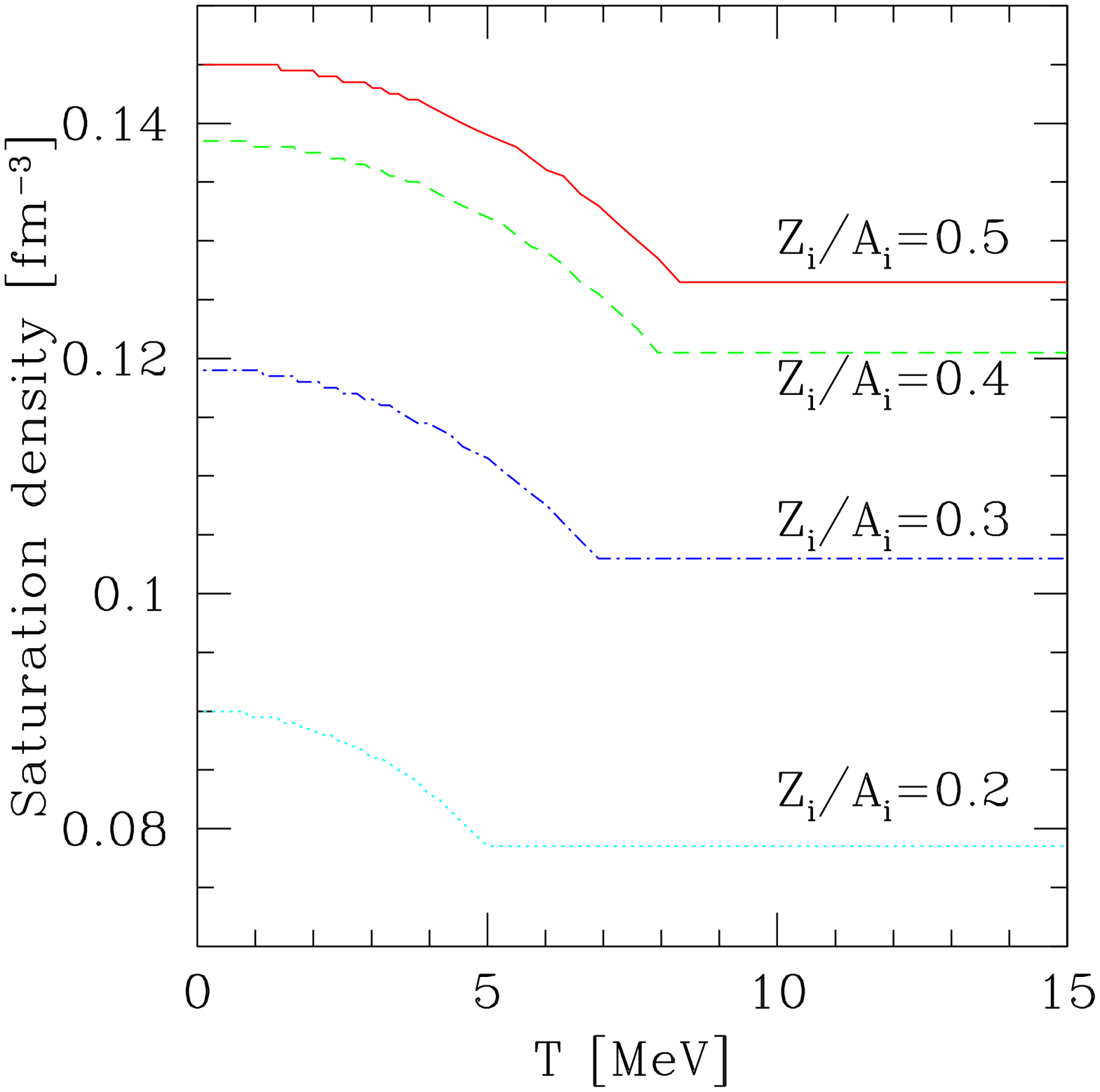}
    \end{tabular}
\caption{The saturation densities of nuclei  as a  function of temperature 
for $Z_i/A_i=$0.2 (cyan dotted line), 0.3 (blue dashed dotted line), 0.4 (green dashed line) and  0.5 (red solid line).  }
    \label{satu}		
        \begin{tabular}{c}
        \includegraphics[width=90mm]{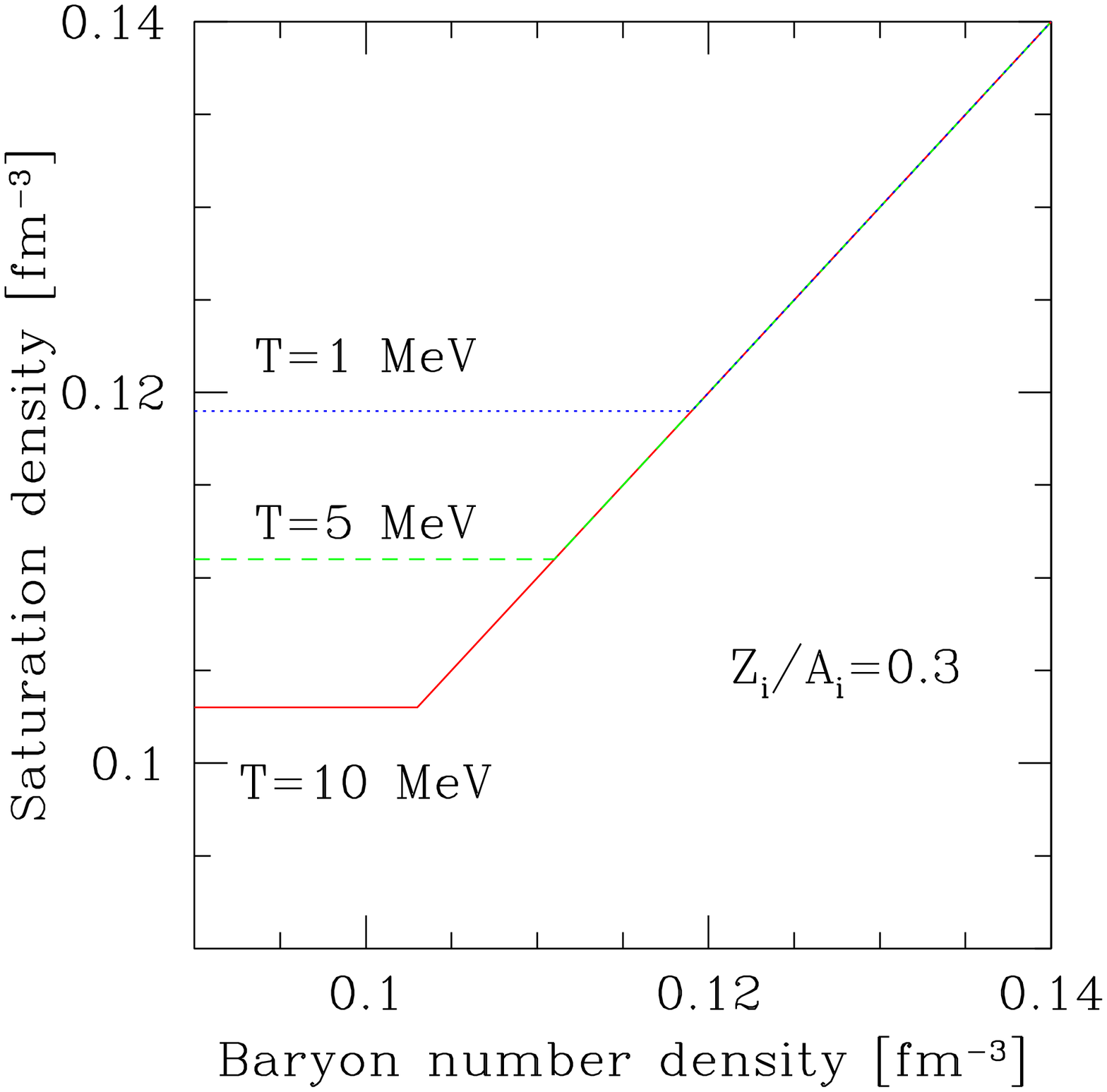}
    \end{tabular}
\caption{The saturation densities of nuclei with  $Z_i/A_i=$0.3 as a function of baryon number density 
for $T=$ 1~MeV (blue dotted line), 5~MeV (green dashed line) and  10~MeV (red solid line).  }
    \label{satu2}		
    \end{center}
\end{figure}

%the saturation densities
We define the saturation densities of nuclei $n_{si}(T)$ 
as the baryon number density, at which the free energy per baryon
 $F^{RMF}(T,n_B, Y_p)$ given by the RMF with $Y_p=Z_i/A_i$ takes its minimum value. 
Thus $n_{si} (T)$ depends on the temperature $T$ and the proton fraction in each nucleus $Z_i/A_i$.
At high temperatures the free energy, $F^{RMF}(T,n_B, Z_i / A_i)$, has no minimum 
because the entropy contribution, the $TS$ term with $S$ being entropy, overwhelms the internal energy.
In the previous paper, $n_{si}$ is set to the saturation density given by H. Shen EOS
 at temperatures higher than the critical temperature $T_{ci}$,
above which the free energy, $F^{RMF}(T,n_B, Z_i / A_i)$, has no minimum. 
This prescription brought unphysical jumps in the mass fraction at the critical temperatures in the previous EOS.
In order to remedy this artifact, we assume in the new EOS
 that the saturation density $n_{si} (T)$ above $T_{ci}$  is equal to the saturation density at the critical temperature $n_{si}(T_{ci})$.
Fig.~\ref{satu} shows the saturation density $n_{si}(T,Z_i/A_i)$  for the proton-fractions $Z_i/A_i = 0.2, 0.3, 0.4$ and $0.5$ in the $n_B-T$ plane. 
We can see that neutron-rich nuclei have lower saturation densities and critical temperatures than symmetric nuclei because of the symmetry energy. 
When the saturation density $n_{si}$ is lower than the baryon number density of the whole system $n_B$, we reset the saturation density as the baryon number density $n_{si}=n_B$
as shown in Fig.~\ref{satu2}.
This prescription approximately represents compressions of nuclei near the saturation densities.
These treatments of the saturation density are important in obtaining reasonable bulk energies
 at high temperatures and densities. 
They are necessary, since it is impossible at the moment to solve nuclear structures and abundance in 
a self-consistent manner completely. 
In fact, the density of each nucleus is not a quantity to be determined by the minimization of
 the free energy density but a parameter to be set in our model. 
We expect, however, that the density of each nucleus is very close to the saturation density, that is, the density, 
at which the free energy density of uniform nuclear matter becomes minimum for the same temperature and 
proton fraction except when the saturation density does not exist at high temperatures or
 when the transition to uniform nuclear matter occurs at a density higher than the saturation density. 
To these cases we need special cares as described above. % rp7

%E_c E_s bulk 23
\subsubsection{Coulomb and surface energies}
To calculate the Coulomb and surface energies of nuclei we set the Wigner-Seitz cell (W-S cell) for each species of 
nuclei so that the charge neutrality could be satisfied. 
Each nucleus is centered in the W-S cell with the volume, $V_i$.
The cell also contains free nucleons as a vapor outside the nucleus as well as electrons, which are assumed to be uniform 
in the entire cell. The charge neutrality in the cell gives the cell volume $V_i = (Z_i - n'_p  V_i^N)/(n_e-n'_p)$ 
where $V_i^N$ is the volume  of the nucleus in the cell and can be calculated as $V_i^N = A_i / n_{si}$. 
The vapor volume and nucleus volume fraction in the cell are given by $ V_i^B = V_i-V_i^N  $ and $u_i =  V_i^N / V_i$, respectively.

In this EOS we assume that each nucleus enters the nuclear pasta phase individually when 
the volume fraction, $u_i$, reaches $0.3$ and that the bubble shape is realized when it exceeds $0.7$ \citep{watanabe2005}. 
The bubbles are explicitly treated as  nuclei of spherical shell shapes with the vapor nucleons filling the inside.
This phase is important to ensure continuous transitions to uniform matter as noted in \citet{Furusawa2011}.
The intermediate states ($ 0.3<u_i<0.7 $) are smoothly interpolated from the normal and bubble states. 
The criterion of intermediate states ($0.3<u_i< 0.7$) is 
admittedly rather arbitrary, although we consulted the literature \citep{watanabe2005} in adopting these numbers.
We have hence tried another choice, $0.4 < u_i < 0.6$, and confirmed that the thermodynamic 
quantities are hardly affected. On the other hand, the nuclear composition is rather sensitive 
to the criterion particularly when the temperature is low and most of nuclei form pastas simultaneously, 
since the surface and Coulomb energies are modified. Since the density region that corresponds to 
the intermediate states is narrow and the sums of Coulomb and surface energies for the drop and bubble
 states are equal to each other at $u_i = 0.5$, the inclusion of the intermediate phase is chiefly meant to ensure
 the smooth change in mass fractions of nuclei around $u_i = 0.5$. %rp8
The evaluation of the Coulomb energy in the W-S cell is given by the integration of Coulomb forces in the cell:
\begin{eqnarray}
E_i^C=\left\{ \begin{array}{ll}
\displaystyle{\frac{3}{5}\left(\frac{3}{4 \pi}\right)^{-1/3}  \frac{e^2}{n_{si}^2} \left(\frac{Z_i - n'_p V_i^N}{A_i}\right)^2 {V_i^N}^{5/3} D(u_i)}    & (u_i\leq 0.3), \\
\displaystyle{\frac{3}{5}\left(\frac{3}{4 \pi}\right)^{-1/3}   \frac{e^2}{n_{si}^2} \left(\frac{Z_i - n'_p V_i^N}{A_i}\right)^2 {V_i^B}^{5/3} D(1-u_i)}   & (u_i\geq 0.7),
\end{array} \right.
\label{eqclen}
\end{eqnarray} 
with $D(u_i)=1-\frac{3}{2}u_i^{1/3}+\frac{1}{2}u_i$, where $e$ is the elementary charge. 

The surface energy of nuclei is given by the product of the nuclear surface area and the surface tension. 
\begin{eqnarray} \label{eq:surf}
E_i^{Su}&=&\left\{ \begin{array}{ll}
4 \pi {r^2_{Ni}} \, \sigma_i \left(1-\displaystyle{\frac{n'_p+n'_n}{n_{si}}} \right)^2 =
4\pi \left( \displaystyle{\frac{3}{4 \pi}} V_i^N \right)^{2/3}  \, \sigma_i  \left(1-\displaystyle{\frac{n'_p+n'_n}{n_{si}}} \right)^2   & (u_i\leq 0.3), \\
4 \pi {r^2_{Bi}} \, \sigma_i  \left(1-\displaystyle{\frac{n'_p+n'_n}{n_{si}}} \right)^2=4\pi \left(\displaystyle{\frac{3}{4 \pi}} V_i^B \right)^{2/3} \, \sigma_i 
\left(1-\displaystyle{\frac{n'_p+n'_n}{n_{si}}} \right)^2  & (u_i\geq 0.7), 
\end{array} \right.\\
&&\sigma_{i}=\sigma_0  - \frac {A_i^{2/3} } {4 \pi r_{i}^2} [S_s(1- 2(Z_i/A_i)^2) ],
\end{eqnarray} 
where  $r_{Ni} = ( 3/4 \pi V^N_i)^{1/3}$ and $r_{Bi} = ( 3/4 \pi V^B_i)^{1/3}$ are the radii of nucleus and bubble.
$\sigma_0$ denotes the surface tension for symmetric nuclei. 
The surface tension  $\sigma_i$  includes the surface symmetry energy,
 i.e., neutron-rich nuclei have lower surface tensions than symmetric nuclei.
 The values of the constants, $\sigma_0=1.15 \rm{MeV/fm^3}$ and 
$S_s =45.8 \rm{MeV}$, are adopted from the paper by \cite{Lattimer1991}. 
The appropriate estimation of surface tensions is important, since they have a critical influence on the abundance of nuclei
 and, as a consequence, on the average mass number of nuclei, as shown in \citet{Buyukcizmeci}.
We may choose other values such as those given in \citet{Lee10},
which include high-order temperature dependences. 
We prefer the simpler estimate by \citet{Lattimer1991} in this work,
 considering insufficient experimental information
 on the heavy and/or neutron-rich nuclei that exist in the supernova matter. %rp9}
The last factor in Eq.(\ref{eq:surf}), $ \left(1-(n'_p+n'_n)/n_{si} \right)^2 $, is
assumed to take into account the effect that the surface energy should be 
reduced as the density contrast decreases between the nucleus and the nucleon vapor.
We use cubic polynomials of $u_i$ for interpolation between the droplet and bubble phases. 
The four coefficients of the polynomials are determined by the condition that the Coulomb and surface energies are continuous and smooth as a function of $u_i$ at $u_i=0.3$ and $u_i=0.7$. 
%shell and bullk  21

\subsubsection{Bulk and shell energies} 
We derive the bulk energies from the free energy per baryon of the uniform nuclear matter
 at the saturation density $n_{si}$ for the given temperature $T$ and proton fraction inside the nuclei $Z_i/A_i$ as
\begin{equation} \label{eq:bulk}
E_{i}^{B} =A_i \{m_B +  F^{RMF}(n_{si},T,Z_i/A_i) \},
\end{equation}
where $F^{RMF}(n_B,T,Y_p)  $ is the free energy per baryon given by the RMF, 
which is the same as that for the free energy density of free nucleons. 
Note that this bulk energy includes the symmetry energy of nuclei. 
In the previous paper, Eq.(7) is applied only to the nuclei with no experimental mass data.
For the nuclei with experimental mass data available, on the other hand, 
the bulk energies are calculated as $E_{i}^{B+sh} =M_i^{data}-[E_i^C+E_i^{Su}]_{vacuum}$ including the nuclear shell energies.
Then  they have no temperature dependence.
In this paper, we evaluate the bulk energies of all heavy nuclei by  Eq.~(\ref{eq:bulk})
so that the bulk energies of all nuclei would  depend on the temperature. 

\begin{figure}
   \begin{center}
         \includegraphics[width=100mm]{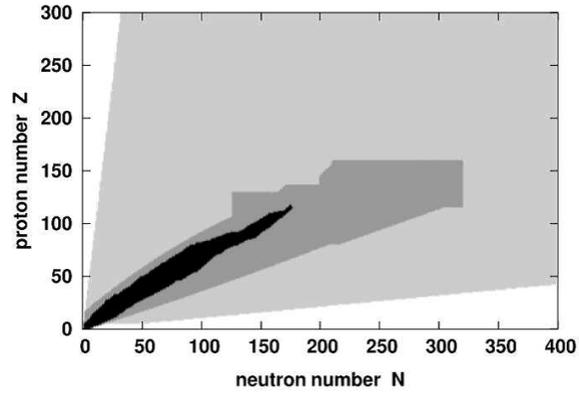}
\caption{The regions of experimental mass data from \citet{Audi2003} (black), theoretical mass data form \citet{Koura2005} (dark gray) 
and the nuclei calculated by LDM with no shell effects (light gray). The upper limits of proton and neutron numbers are 1000.}
    \label{nc}		
    \end{center}
\end{figure}

We include the shell effects separately in the mass formula of nuclei by using both experimental and
 theoretical mass data \citep{Audi2003, Koura2005} to better reproduce the ordinary NSE EOS results in the low density regime.
The regions, in which the experimental and theoretical mass data are available, are shown in nuclear chart in the Fig.~\ref{nc}.
The shell energies are obtained from the experimental or theoretical mass data  by subtracting our liquid drop mass formula,
which does not include the shell effects, $(M_i^{LDM}=E_i^B+E_i^C+E_i^{Su})$ in the vacuum limit as $E_{i}^{Sh} =M_i^{data} -[ M_i^{LDM} ]_{vacuum}$.
The vacuum limit means that the nucleus is cold and isolated: $T, n'_{p/n}, n_e =0$.
At high densities, the shell effect of nuclei estimated in vacuum is
 considered to be diminished because of the existence of electrons, free nucleons and other nuclei.
We take this effect into account phenomenologically as follows:
\begin{eqnarray}\label{eq:sh}
   E_{i}^{Sh}&=&\left\{ \begin{array}{ll}
M_i^{data}-[E_i^B+E_i^C+E_i^{Su}]_{vacuum} & (\rho \leq 10^{12} \rm{g/cm^3}), \\
(M_i^{data}-[E_i^B+E_i^C+E_i^{Su}]_{vacuum}) \\
\ \ \ \ \times (\rho_0- \rho) /(\rho_0 - 10^{12}  \rm{g/cm^3})& (\rho >  10^{12} \rm{g/cm^3}), 
\end{array} \right.
\end{eqnarray} 
where $\rho_0$ is taken to be $m_B$ times  the saturation density of symmetric nuclei $n_{si}(T, Z_i/A_i=0.5)$ at temperature $T$.
The last factor $(\rho_0- \rho)/(\rho_0 - 10^{12} \rm{g/cm^3})$ accounts for the decay of shell effects at high densities. 
The choice  of the critical density $10^{12} \rm{g/cm^3}$  is rather arbitrary, 
since the dependence of shell energies on the density of ambient matter has not been thoroughly investigated yet.
It is noted, however, the structure of nuclei is known to be affected by ambient matter at these densities.
The abundances of nuclei with magic numbers of protons or neutrons are affected by the shell energy and hence by the choice of the critical density. 
We have confirmed, however, that thermodynamics quantities are hardly changed for the critical density of  $10^{13} \rm{g/cm^3}$. %rp10
The linear interpolation in Eq.~(\ref{eq:sh}) makes the free energy not smooth and the pressure discontinuous at the boundaries of the interpolation region. 
In practice, however, the variation of the shell energy is quite minor compared with those of Coulomb and translational 
energies and the discontinuities of the pressure are negligible.

We neglect the shell energies of the heavy or neutron-rich nuclei with no available mass data,
since we have no guidance to estimate the shell energy and 
such nuclei are abundant only at very high densities, where the shell effects will be minor anyway.

We sum up all the contributions to have the  masses of heavy nuclei as $M_{i} =E_i^B +E_i^{Sh} + E_i^{C} +E_i^{Su}$.
For the nuclei with mass data available, this formula can be transformed to
\begin{equation}
  M_{i}=M_i^{data} + \Delta E_i^{B} +\Delta E_i^C +\Delta E_i^{Su} \ \ \ \  (\rho \leq 10^{12} \rm{g/cm^3}), 
\end{equation}
where $\Delta E_i$ means the difference from the vacuum limit: $\Delta E_i= E_i-[E_i]_{vacuum}$. 
In the limit of low densities and temperatures, $M_{i}$ is reduced to the mass data $M_i^{data}$.
% because $\Delta E_i^{B}$, $\Delta E_i^C$ and $\Delta E_i^{Su}$  are small.
This feature is important for reproducing the ordinary NSE results in these limits \citep{Timmes1999}.
At the saturation density, on the other hand, only the bulk energies $E_i^B$ survive,
 since other terms are diminished as the density approaches the saturation density in our model. 

%%%%%%%%%%%%%%%%%%%%%%%%%%%%%%%%%%%%%%%%%%%%%%%%%%%%%%%%%%%%%%%%%%%%%%%%%
\subsection{Mass evaluation of light nuclei ($Z\leq 5$) \label{secln}} 
In this subsection, we explain how to evaluate the masses of light nuclei ($Z \leq 5$). 
Note that the mass  formula employed for heavy nuclei, which is based on LDM, is inappropriate for light nuclei as already noted.
We assume  the descriptions of $d,t,h$ and $\alpha$  in dense and hot matter based on quasi-particles outside heavy nuclei 
and no pasta phase is considered for them.
The saturation densities of the four light nuclei are set to the constant value, 0.15 $\rm{fm}^{-3}$,
 in contrast to those of heavy nuclei, which depend on temperatures and densities. 
For the light nuclei $(Z\leq5)$  other than $d, t, h$ and $\alpha$ such as  $ ^7\rm{Li}$, we adopt the mass data 
with density and temperature corrections that are based on the LDM slightly different from that of heavy nuclei (see below for details). 

The masses of  $d, t, h$ and $\alpha$  are given by the following expression:
\begin{equation}
  M_{j}=M_j^{data} + \Delta E_{j}^{Pa}  + \Delta E_j^{SE} +\Delta E_j^C \ \ \ \ (j=d,t,h \  \&  \ \alpha) 
\end{equation}
where $\Delta E_j^{Pa}$ is the Pauli energy shift by other baryons, 
$\Delta E_j^{SE}$ is the self-energy shift of the nucleons composing the light nuclei and  $\Delta E_j^C$ is the coulomb energy shift.

For the Pauli energy shifts of the light nuclei, we employ the empirical formulae provided by \citet{Typel2010},
 which are quadratic functions fitted to the result of quantum statistical calculations \citep{Ropke2009}.
R{\"o}pke investigated the binding energies of light clusters 
in hot and dense matter $(T \lesssim 20$~MeV and $n_B \lesssim 0.16 \ \rm{fm^{-3}})$ 
by using the quantum statistical approach.
They regard the light clusters $d, t, h$ and $\alpha$ as quasi-particles and
 solve the in-medium Sch{\"o}rdinger equation perturbatively.
For the potential terms in pair interactions, Jastrow and Gaussian wave-function approximations are adopted for $d$ and others $(t, h,\alpha)$, respectively.
%, which is a successful model used to describe nuclear structure in a wide region of $(N,Z)$ plane
%and has also been tested in heavy-ion collisions.
Note that the fitting formulae of $\Delta E_j^{Pa}$  in \citet{Ropke2009} are obtained 
under the assumption that matter is composed of only nucleons and light clusters $(Z \leq 2 $ and $N \leq 2)$,
which is not completely consistent with the situations of our interest, in which heavier nuclei are also existent.
To obtain the Pauli energy shifts, we define the local proton and neutron number densities including light nuclei as:
\begin{eqnarray}
n_{pl}=n'_p + \eta^{-1} \sum_{j=d,t,h,\alpha} {Z_j n_j}  \\
n_{nl}=n'_n + \eta^{-1} \sum_{j=d,t,h,\alpha} {N_j n_j} \ . 
\end{eqnarray}
where $\eta$ stands for the volume fraction $(V-V_N)/V$. 
%In calculation of Pauli energies, we use substitute the total proton $n_B Y_p$ and neutron densities $n_B (1-Y_p)$ 
%for the local number densities $n_{pl}$ and $n_{nl}$ from the viewpoint that
% all baryon, which are composed of not only light clusters but heavy nuclei, contribute the Pauli energy shifts of light nuclei.
%On the other hands,we use $n_{pl}$ and $n_{nl}$ for self-energy shifts of nucleons compounded in light nuclei,
%since we assume the self-energies of light nuclei depend on  local densities as same as free nucleons. 
Then the Pauli energy shift $\Delta E_j^{Pa}$ is given by the following expression:
\begin{equation}
\label{DEP}
 \Delta E_{j}^{Pa}(n_{pl},n_{nl},T) = - \tilde{n}_{j} 
 \left[1 + \frac{\tilde{n}_{j}}{2\tilde{n}_{j}^{0}(T)} \right]
 \delta B_{j} (T),
\end{equation}
which is quadratic in $\tilde{n}_{j} = 2(Z_{j} \ n_{pl} +N_{j} \ n_{nl}) /A_{j}$.
The density scale for the dissolution of each light nucleus is given by  $\tilde{n}_{j}^{0}(T) = B_{j}^{0}/\delta B_{j}(T)$
with the binding energy in vacuum, $B_{j}^{0}=Z_j M_p+N_j M_n -M_j^{data}$. 
The function $\delta B_{j}(T)$  represents the temperature dependence of the Pauli energy shifts
and is originally derived with the Jastrow and Gaussian wave-function approximations for $d$ and other light nuclei, respectively, as %rp11
\begin{eqnarray}
   \delta B_{j}(T) &=&\left\{ \begin{array}{ll}
  a_{j,1} / T^{3/2}  \left[ 1/\sqrt{y_{j}}  -  \sqrt{\pi}a_{j,3} \exp  \left( a_{j,3}^{2} y_{j} \right) 
 {\rm erfc} \left(a_{j,3} \sqrt{y_{j}} \right) \right]   &  {\rm{for}} \   j  =  d, \\
  a_{j,1} /\left( T  y_j \right)^{3/2}  \ \   &  {\rm{for}} \  j  =  t,h,\alpha,
\end{array} \right.
\end{eqnarray} 
with $y_{j} = 1+a_{j,2}/T$.
The parameters $a_{j/1}$ $a_{j/2}$and  $a_{j/3}$ are given in Table~\ref{tab1}. 

The self-energy shifts of light nuclei are the sum of
 the self-energy shifts of individual nucleons composing the light nuclei $E^{SE}_{n/p}$ and 
the contribution from their effective masses $\Delta E_{j}^{\rm eff.mass}$:
\begin{equation}
\label{SE}
\Delta E_{j}^{SE}(n'_{p},n'_{n},T)= (A_j-Z_j) \Delta E_{n}^{SE}+ Z_j \Delta E_{p}^{SE} +\Delta E_{j}^{\rm eff.mass}\ 
\end{equation}
where $\Delta E^{SE}_{n/p}=\Sigma^0_{n/p}(T,n'_p,n'_n)-\Sigma_{n/p}(T,n'_p,n'_n)$ 
with $\Sigma^0$ and $\Sigma$ being the vector and scalar potentials of nucleons.
The effective mass contributions are given as
$\Delta E_{j}^{\rm eff.mass} = \left(1-m^{\ast}/m \right)s_{j}$
with $m^{\ast}=m_B-\Sigma_{n/p}(T,n'_p,n'_n)$.
The coefficients $s_j$ for $d,t,h,\alpha$ are given in table \ref{tab1}.
The potentials $\Sigma^0$ and $\Sigma$ are calculated from the RMF employed for free nucleons in this paper.
Note that the coefficients  $s_j$ are provided based on a different RMF theory with density-dependent meson-nucleon couplings \citep{Typel2005}.  
However, the inconsistency should have little influence, since the effective mass term is in general smaller than the other two potential terms
and the light nuclei are not abundant  at high densities, where the effective mass terms could be large, due to the Pauli energy shifts
 and pasta formations of heavy nuclei.
 %and provided as a function of number density $n$, asymmetry $\delta$ and temperature $T$ in Appendix A of \citet{Typel}.
%As previously noted, we calculate the potentials with $n=n_{pl}+n_{nl}$ and $\delta=(n_{pl}-n_{nl})/(n_{pl}+n_{nl})$.

More detailed explanations of the Pauli- and self-  energy shifts are provided in \citet{Typel2010}.
Note that we neglect the dependence of the Pauli energy shifts on the momentum of the light clusters and 
that of the self-energy shifts on the momentum of nucleons composing the light clusters for simplicity.
%In the NSE description, the momentum dependence of nuclear mass energies can not be calculated easily and we use the mass energies of light nuclei with zero momentum. 
 
The Coulomb energy shifts are calculated as
\begin{eqnarray}
 \Delta E_j^{C} &= &E_j^C(n'_p,u_j)-E_j^C(0,0), \\ 
 E_j^C (n'_p,u_j) &= &\frac{3}{5}\left(\frac{3}{4 \pi}\right)^{-1/3}  \frac{e^2}{n_{sj}^2} \left(\frac{Z_j - n'_p V_j^N}{A_j}\right)^2 {V_j^N}^{5/3} D(u_j). 
\end{eqnarray} 
Although the evaluation of the Coulomb energy is identical to that for heavy nuclei
 in the droplet phase, the shifts are negligible  compared with other energies.
We do not take into account the nuclear pasta phases
 and surface energy shifts for the light nuclei.
 
The light nuclei $(Z_j \leq 5)$ other than $d, t, h$ and $\alpha$ are described by an LDM.
Since the masses of light nuclei $d, t, h$ and $\alpha$ are almost unchanged at low densities,
the Pauli- and self-energy shifts are negligible at low densities.   
Therefore we assume that the temperature dependence of other light nuclei is not so strong at low densities either
and the temperature dependence is important only at $\rho >  10^{12} \rm{g/cm^3}$, which are approximated as 
\begin{eqnarray}
\Delta M_{j}= & \left\{ \begin{array}{ll}
 M_j^{data} + \Delta E_j^{Su}+\Delta E_j^{C} & (\rho \leq 10^{12} \rm{g/cm^3}), \\
 M_j^{data} + \Delta E_j^{Su}+\Delta E_j^{C} % \times\frac{{\displaystyle (\rho_0 - \rho)}}{{\displaystyle (\rho_0 - 10^{12}\rm{g/cm^3})}}  \\ 
 + \{ A_j F_j^{RMF}(n_{sj},T,Z_j/A_j) & \\
 -(M_j^{data} - [E_j^{Su}+E_j^{C}]_{vacuum}) \} 
\times\frac{{\displaystyle (\rho - 10^{12}\rm{g/cm^3})}}{{\displaystyle (\rho_0 - 10^{12}\rm{g/cm^3})}} 
 & (\rho >  10^{12} \rm{g/cm^3}), 
\end{array} \right.
\end{eqnarray}
where $\Delta E_j^{Su}$ is the surface energy shift, which is too small to make any difference except in the pasta phases.
The self-energy shift is linearly interpolated between the bulk and shell energies in vacuum limit,
which are estimated from the experimental mass data by subtracting surface and Coulomb energies in vacuum limit,
and the self-energy of uniform matter obtained by the RMF theory.
The Pauli energy shifts are neglected for these light nuclei,
 since no fitting formula is available.
We assume that they experience the  pasta phases  in the same way as heavy nuclei.
The Coulomb and surface energy shifts are calculated from the same LDM for heavy nuclei.
We note that the light nuclei other than $d,t,h $ and $ \alpha$ are not so important because they are never  abundant under NSE,
 since $d, t, h$ and $\alpha$ are dominant over the other light nuclei at high temperatures and/or low densities,
 whereas heavy nuclei 
prevail in the opposite situations. 

In our models, $d,t,h$ and $\alpha$ are treated as independent particles and they coexist with free nucleons outside heavy nuclei.
%In contrast, the other nuclei are described by the liquid drop model. 
%They are connected with each other at high densities.
At low densities, the masses of the light nuclei approach  the experimentally known values,
 since the $\Delta E_j^{Pa}$, $\Delta E_j^{SE}$ and $\Delta E_j^C$ vanish in this limit.
Near the saturation densities, light nuclei no longer exist because of the Pauli energy shifts  
and free nucleons and heavy nuclei in the pasta phases are abundant. 
%%%%%%%%%%%%%%%%%%5%%%%%%%%%%
\subsection{Translational energies of nuclei} \label{sectr}
% E_t   
The translational energy of nucleus $i$ in our model free energy is based on that for the ideal Maxwell-Boltzmann gas and given by 
\begin{equation}
\label{eq:tra}
 F_i^{t}=k_B T \left\{ \log \left(\frac{n_i}{g_i^0 n_{Qi}}\right)- 1 \right\} \left(1-\frac{n_B}{n_s}\right), 
\end{equation}
where $k_B$ is the Boltzmann constant and $n_{Qi} = \left(M_{i/j} k_B T/2\pi \hbar ^2  \right)^{3/2}$, 
$g_i^0$ is the spin degree of freedom of the ground state.
Note that the contribution of the  excited states to free energy is encapsulated in the temperature dependence of the bulk energy. 
In the previous paper, we employed a functional form of  $g_i(T)$ for the internal degree of freedom in Eq.~(\ref{eq:tra}). 
%to take into account the contribution of excited states.
The last factor on the right hand side of Eq.~(\ref{eq:tra}) takes account of the excluded-volume effect: each nucleus can move in 
the space that is not occupied by other nuclei and free nucleons. The factor reduces the translational energy at high densities and
is important to ensure the continuous transition to uniform nuclear matter.
The present form of the factor, $(1-n_B/n_s)=(V-V_{baryon})/V$, gives a linear suppression in terms of the occupied volume $V_{baryon}$
and we always employ the nuclear saturation density for symmetric nuclei $n_s=\left[ n_{si}(Z_i/A_i,T) \right] _{Z_i/A_i=0.5}$ for numerical convenience.
%%%%%%%%%%%%%%%%%%%%%%%%%%%%%%%%%%%%%%%%%%%55
%%%%%%%%%%%%%%%%%%%%%%%%%%%%%%%%%%%%%%%%%%%%%%%%%%%%%%%%%%%%%%%%%%%%%%%%%%%%%%%%%%%%%
\subsection{Thermodynamical quantities \label{secth}}%
After minimization, we obtain the free energy density together with the abundances of all nuclei 
and free nucleons as a function of $\rho_B$, $T$ and $Y_p$. 
Other physical quantities are derived by partial differentiations of the free energy density.
In  so doing, all the terms concerning the excluded volume effects and the interpolation factors
are properly taken into account to ensure the thermodynamical consistency as described in \citet{Furusawa2011} in detail. %rp6
The baryonic pressure, for example, is obtained by the differentiation with respect to the baryonic density
as follows:  
\begin{eqnarray}
 p_B &=&n_B \left[\partial{f}/\partial {n_B}\right]_{T,Ye}-f,  \nonumber \\
   &=&   p_{p,n}^{RMF}+ \sum_{i/j\neq d,t,h,\alpha} ( p_{i/j}^{th}  + p_{i/j}^{ex}  + p_{i/j}^{mass}), \label{eq:pr} \\ 
   p_{i}^{mass}&= & (p_{i}^{shell} + p_{i}^{Coul} + p_{i}^{Surf}) \ \ \ \  \ \ \  (\rm{heavy  \ nuclei} \ \it{Z_i} \geq \rm{6}), \\ 
   p_{j}^{mass}&= & (p_j^{Pauli} + p_j^{SE} + p_j^{Coul})  \ \ \ \ \  \  \ \ \ \ (d, t, h, \alpha),    \label{eq:prl} \\
   p_{j}^{mass}&= & ( p_{j}^{SE} + p_{j}^{Coul}+p_{j}^{Surf}) \ \ \ \ \ \ (\rm{other \  light \ nuclei} \ \it{Z_j} \leq \rm{5}),
\end{eqnarray}
where $ p_{p,n}^{RMF}$ is the contribution of the nucleons in the vapor; 
both $p_{i/j}^{th}$  and $p_{i/j}^{ex}$ come from the translational energy of nuclei in the free energy;
$p_{i/j}^{shell}$, $p_{i/j}^{Coul}$ and $ p_{i/j}^{Surf}$ 
originate from the shell, Coulomb and surface energies of nuclei in the free energy, respectively;
 $p_j^{Pauli}$ and $p_j^{SE}$ are derived from the Pauli- and self-energy shifts of the light nuclei.

The entropy per baryon is calculated from the following expression:
\begin{eqnarray}
 s& =& -\frac{\left[\partial{f}/\partial {T}\right]_{\rho_B,Ye}}{n_B},  \label{eq:ent} \\
    &=&\eta s^{RMF}_{p,n} +  \sum_{i,j} \frac{n_{i,j} k_B}{n_B} \Biggl[ \left\{  \frac{5}{2}- \log \left(\frac{n_{i,j}}{g^0_{i,j} n_{Qi}} \right)
   \right\} (1-n_B/n_s) \nonumber  -  \frac{\partial{M_{i,j}}}{\partial {T}}  \Biggl]
\end{eqnarray}
This form is the same as  that of the previous EOS.
The partial derivative of the masses, $\partial{M_{i,j}}/\partial {T}$,  is originated from the temperature-dependence of nuclear mass in the current formulation and given as follows: 
\begin{eqnarray}
\label{eq:ent2}
 \frac{\partial{M_i}}{\partial {T}} =  & \frac{\displaystyle \partial{E_i^{B}}}{\displaystyle \partial {T}} = - A_i s_i^{RMF} (T,n_{si}, Z_i / A_i)  \ \ \ \ \  \ \  
\ \ \ \ \ \ \ \ \ \ \ \ \ \ \ \ \  \ \ \ \ \ \ \ \ \ \ \ (\rm{heavy  \ nuclei}), & \\
 \frac{\partial{M_j}}{\partial {T}} = & \frac{\displaystyle \partial{\Delta E_j^{SE}}}{\displaystyle \partial {T}} + 
\frac{ \displaystyle \partial{\Delta E_j^{Pa}}}{ \displaystyle \partial {T}} \ \ \ \ \ \ \ \ \ \ \ \  \
 \ \ \ \ \ \ \ \ \ \ \ \  \ \ \ \ \ \ \ \ \ \ \ \ \  \ \ \ \ \ \ \ \ \ \ \ \ \  (j=d, t, h, \alpha), &\\
 \frac{\partial{M_j}}{\partial {T}} = & \left\{ \begin{array}{ll}
0 & (\rho \leq 10^{12} \rm{g/cm^3}), \\
 \frac{\displaystyle A_j s_j^{RMF}(n_{sj},T,Z_j/A_j)}{\displaystyle (\rho_0 - 10^{12})} 
 (\rho- 10^{12}) & (\rho >  10^{12} \rm{g/cm^3})    \ \ \  (\rm{other \  light  \ nuclei}),
\end{array} \right.
\end{eqnarray}
where the entropy per baryon $s_{i/j}^{RMF}$ is predicted by the RMF.
The contribution of this term is normally negligible except near the nuclear saturation density.
%
%%%%%%%%%%%%%%%%%%%%%%%%%%%%%%%%%%%%%%%%%%%%%%%%%%%%%%%%%%%%%%%%%%%%%%%%%%%%%%%%%%%%%%%%%%
%%%%%%%%%%%%%%%%%%%%%%%%%%%%%%%%%%%%%%%%%%%%%%%%%%%%%%%%%%%%%%%%%%%%%%%%%%%%%%%%%%%%%%%%%%
%%%%%%%%%%%%%%%%%%%%%%%%%%%%%%%%%%%%%%%%%%%%%%%%%%%%%%%%%%%%%%%%%%%%%%%%%%%%%%%%%%%%%%%%%
\section{Result} 
In this paper, we construct the EOS modifying the previous one \citep{Furusawa2011}.
First we focus on  the changes for heavy nuclei, i.e.
 the employment of the theoretical  mass data and
 the modification of the temperature dependences of the bulk energies and the internal degrees of freedom. 
Then we compare the results of the different modelings of the light nuclei $Z \leq 5$.
We list five calculated models in Table 2. 

Model~0a is nothing but the previous EOS except for the assumption on the saturation densities:
in the previous one, the saturation densities at high temperature are determined by H. Shen EOS
whereas in the new models, they are derived from the RMF calculation as noted in section \ref{sechn}.
In Model~0a  the temperature dependence of the bulk energies of the nuclei, for which the mass data are available, is neglected
and the bulk energies including the shell energies are derived from 
the experimental mass minus the Coulomb and surface energies in vacuum as
\begin{eqnarray}
  & & E_{i}^{B+Sh}=  \nonumber\\
& & \left\{ \begin{array}{ll}
 M_i^{data} -[E_i^C+E_i^{Su}]_{vacuum} & (\rho \leq 10^{12} \rm{g/cm^3}), \\
\\
\frac{\displaystyle \{(M_i^{data}-[E_i^C+E_i^{Su}]_{vacuum}) (\rho_0- \rho) +(A_i F_i^{RMF})(\rho - 10^{12} \rm{g/cm^3}) \}}
{\displaystyle(\rho_0 - 10^{12} \rm{g/cm^3})}& (\rho >  10^{12} \rm{g/cm^3}).
\end{array} \right.
\end{eqnarray} 
To take into account the excited states at high temperatures, 
the temperature dependence is introduced in $g_i^0$ in Eq.~(\ref{eq:tra}). % 
The functional form is adopted from \citet {Fai1982} as
\begin{equation}
\label{eq:ex}
g_i(T)=g_i^0 +\frac{c_1}{A_i^{5/3}}\int_0^{\infty} dE e^{-E/T}\exp\left(\sqrt{2 a(A_i) E}\right), \
\end{equation}
in which  $a(A_i)=(A_i/8)(1-c_2 A_i^{-1/3})$~MeV $^{-1}$, $c_1=0.2$~MeV $^{-1}$ and $c_2=0.8$.
More details about the bulk and shell energies $E_{i}^{B+Sh}$ as well as $g_i(T)$ are found in \citet{Furusawa2011}.

In the new Models 1a, 2a, 2b and 2c, we employ the temperature-dependent bulk energies $E_i^{B} (T)$ in Eq.~(\ref{eq:bulk}) 
instead of Eq.~(\ref{eq:ex}).
The difference between the two treatments is most clearly presented as follows: 
the number density $n_{i}$ of heavy nucleus $i$ depends on the internal degree of freedom $g_i(T)$ and the mass energy $M_i(T)$ as
 $n_i \propto g_i(T) \exp(-M_i(0)/T)$ in Model~0a and 
$n_i \propto g_i(0) \exp(-M_i(T)/T)$ in the other Models 1a, 2a, 2b and 2c.
This leads us to introduce   the effective internal degree of freedom $g^{*}_i(T)=g_i^0 \exp(-(E^{B}_i(T)-E^{B}_i(0))/T)$ 
to express the number density as  $n_i \propto g^{*}_i(T) \exp(-M_i(0)/T)$.
This $g^*_i(T)$ is in general much larger than $g_i(T)$. 
In fact the ratio, $g^{*}_i(T)/g_i(T)$, for $^{56}$Fe is  9.18, 75.9 and 130  at $T=1,5$ and 10~MeV, respectively.  %rp12
For the nuclei with no available experimental mass data in Model~0a, the bulk energies are calculated from Eq.~(\ref{eq:bulk}) 
and, as a result, both the bulk energies $E_i^{B}(T)$ and the internal degrees of freedom $g_i(T)$ depend on the temperature.
This double count of the excited states
leads to the overestimation of abundances of this type of nuclei as shown later. %rp13

The bulk energies in Model~1a are all derived from the RMF calculations with  the temperature dependence included as described in subsection~\ref{sechn}.
We consider that Model~2a is the most realistic model for heavy nuclei.
We utilize the theoretical mass data by \citet{Koura2005} in addition to the experimental mass data in the calculation of the shell energies. 
Models 0a and 1a include only experimental mass data by \citet{Audi2003}, on the other hand. 
Note that we neglect the shell energies for the nuclei with no mass data available
 and that the maximum of proton and neutron numbers are set to $1000$ in all models.

In Models 0a, 1a and 2a, the binding energies of light nuclei are evaluated from the LDM employed for heavy nuclei.
Model~2b  is modified from Model~2a only in the mass evaluation of the light nuclei with $Z \leq 5$. 
The LDM for heavy nuclei and mass data are  identical to those in  Model~2a. 
The masses of the light nuclei in Model~2b are based on the quantum  approach, the details of which are given in subsection \ref{secln}.
To examine the effect of the Pauli- and self-energy shifts, we prepare Model~2c,
in which they are set to $\Delta E^{Pa}=\Delta E^{SE}=0$.
This means that the masses of $d,t,h$ and  $\alpha$ are evaluated as $M_j=M_j^{data}+\Delta E_j^{C}$
in Model~2c. 
We consider that Model~2b is the best among all five models. 
%%%%%%%%%%%%%%%%%%%%%%%%%%%%%%%%%%%%%%%%%%%%%%%%%%%%%
\subsection{Abundances of heavy nuclei}  % 3
\begin{figure}
   \begin{center}
    \begin{tabular}{c}
         \includegraphics[width=95mm,angle=-90]{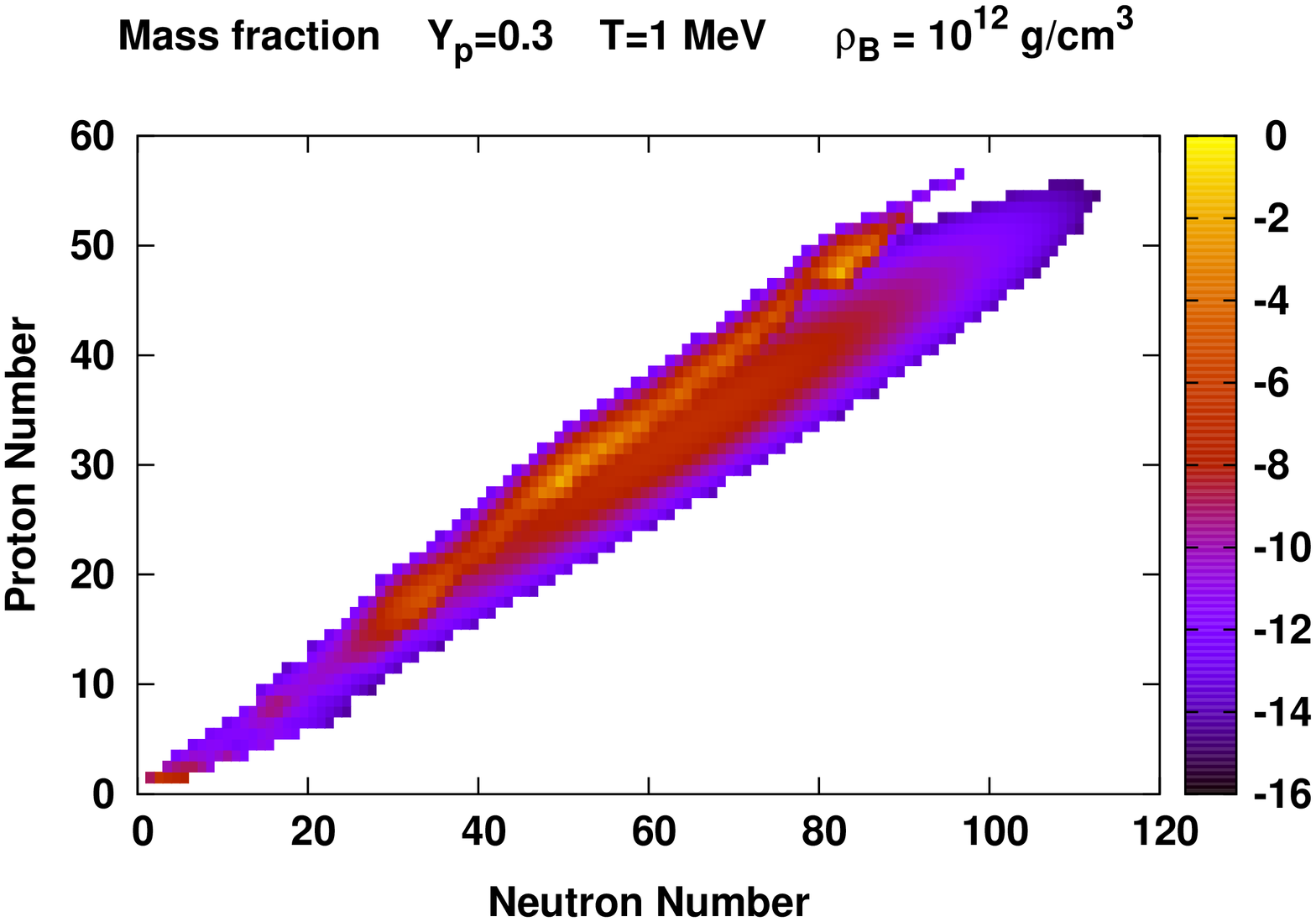}
    \end{tabular}
\caption{The mass fractions in log$_{10}$ of nuclei  in the $(N, Z)$ plane for Model~0a at $\rho_B=10^{12}$g/cm$^3$, $T=1$MeV and $Y_p=0.3$.  }
    \label{dis0}		
        \begin{tabular}{c}
         \includegraphics[width=95mm,angle=-90]{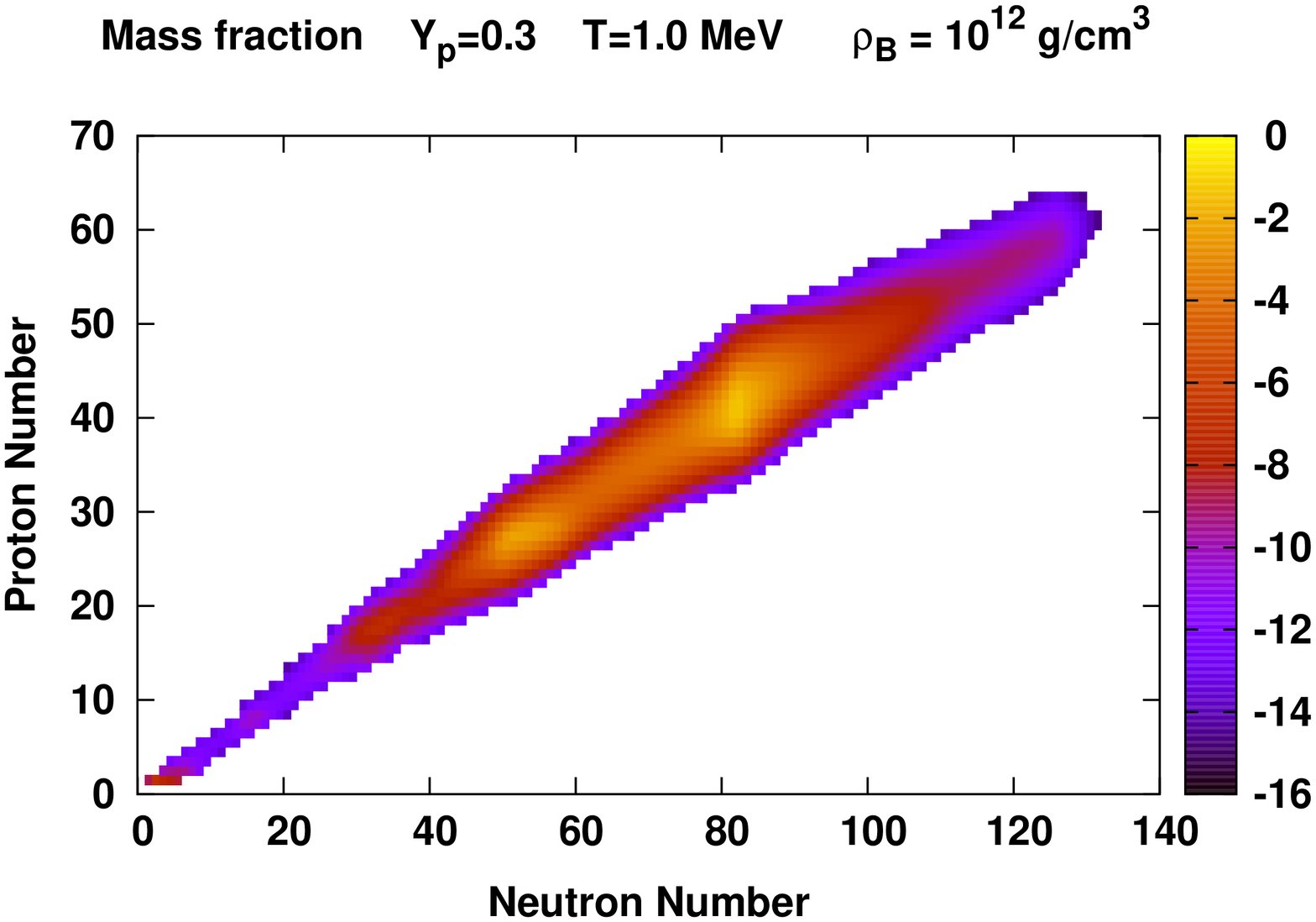}
    \end{tabular}
\caption{The mass fractions  in log$_{10}$  of nuclei in the $(N, Z)$ plane  for Model~2a at $\rho_B=10^{12} $ g/cm$^3$, $T=1$MeV and $Y_p=0.3$. }
    \label{dis1}		
   \end{center}
\end{figure}

%\subsubsection{distributions of nuclei} %9/4
The mass fractions of nuclei for Models 0a and 2a are shown  in the $(N, Z)$ plane for $\rho_B = 10^{12}$g/cm$^3$, $T=1$~MeV and $Y_p=0.3$ 
in Figs.~\ref{dis0} and~\ref{dis1}. 
We can clearly see the gap between the nuclei with the experimental mass data and those without them in Fig.~\ref{dis0}. 
%We can see the discrete abundances especially near the neutron magic number $N=50$. 
On the other hand, there is no such gap in Fig.~\ref{dis1}, since the range of the nuclei,
 for which the shell energies are included, is expanded to  $N \sim 200$ by the use of  theoretical mass data.  
In  Model~2a the mass fractions of the nuclei
 in the vicinities of the magic numbers $N =$ 50 and 82 are enhanced and, as a result,
 the fractions of other nuclei, in particular those with $N \sim 20$, in 2a are smaller.
This effect can be seen also in the isotope abundance discussed in the next paragraph. %rp13

\begin{figure}
   \begin{center}
    \begin{tabular}{c}
               \resizebox{83mm}{!}{\plotone{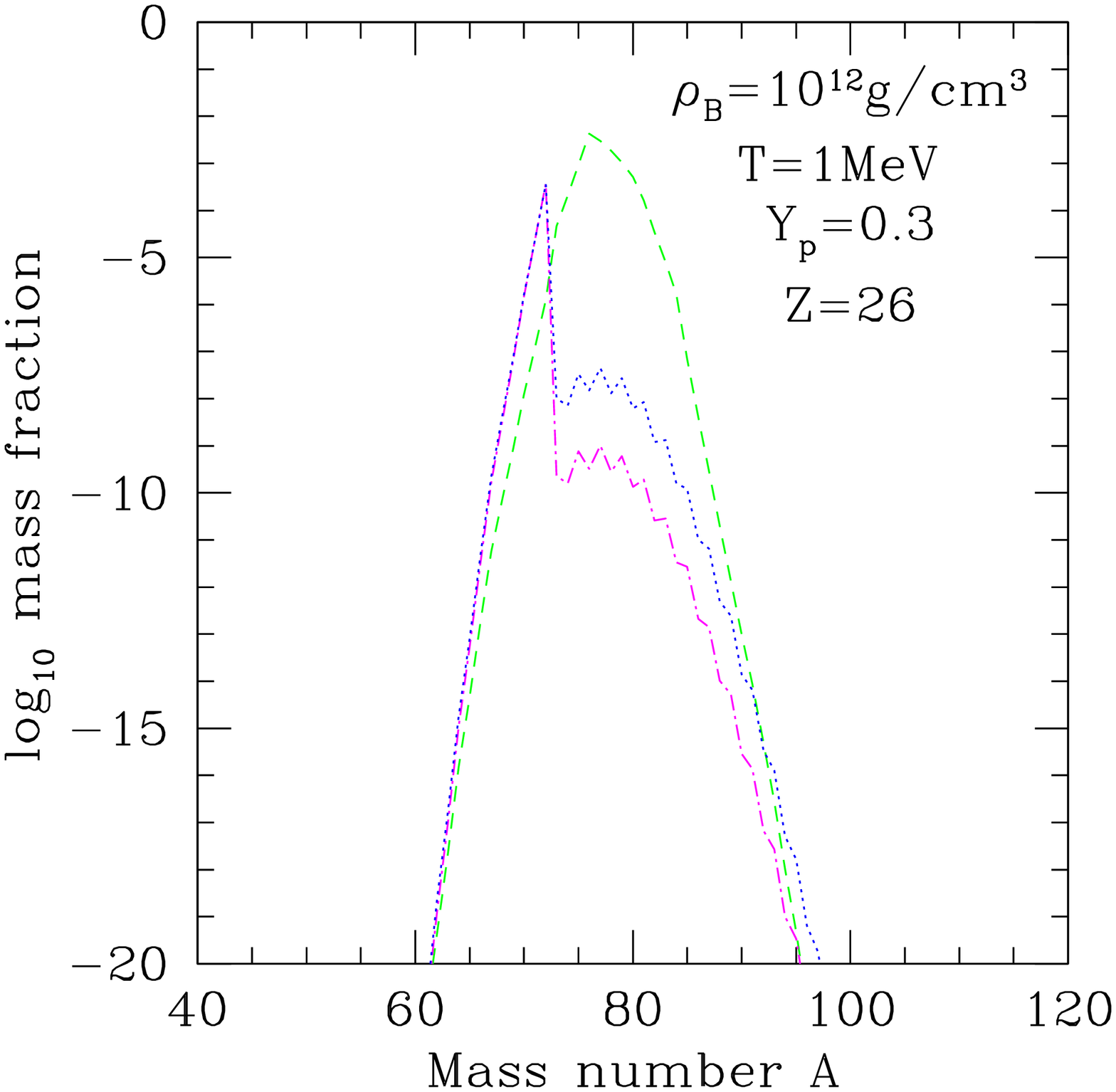}} \\
               \resizebox{83mm}{!}{\plotone{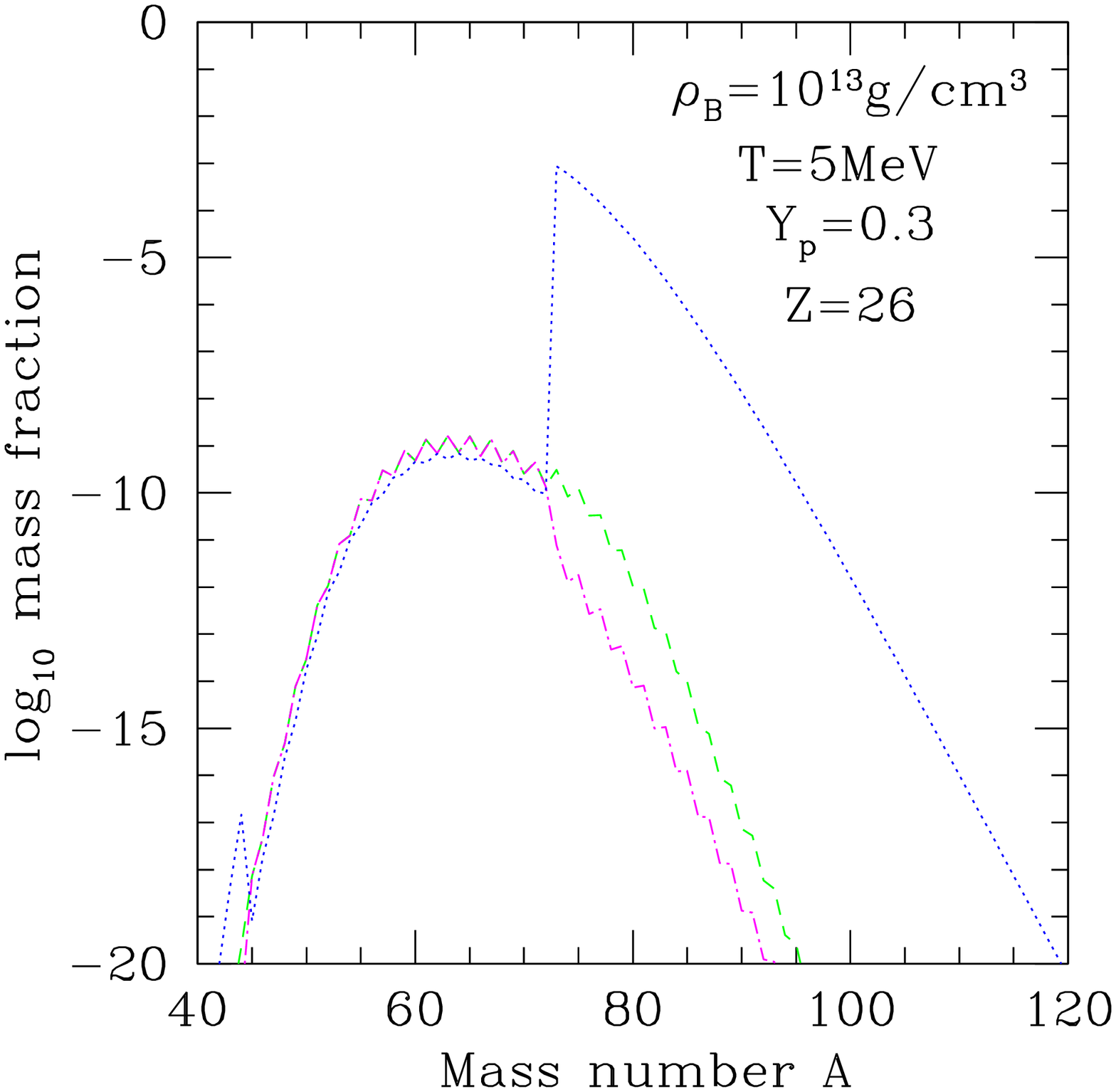}} \\
    \end{tabular}	
   \end{center}
\caption{The isotope abundance of the nuclei with proton number $Z=26$ for Models 0a (blue dotted lines), 1a (magenta dashed dotted lines) and 2a (green dashed lines)  
at  ($\rho_B= 10^{12}$g/cm$^3$, $T=1$MeV, $Y_p=0.3$) and ($\rho_B= 10^{13}$g/cm$^3$, $T=5$MeV, $Y_p=0.3$).}
\label{iso}
\end{figure}

The isotope abundances of the nuclei with $Z=26$ are shown 
for two combinations of temperature and baryon density in Fig.~\ref{iso}. 
The main difference between Model~2a and the others manifests in the range of $ 74 \leq A \leq  99$.
For the nuclei with $Z=26$, the experimental mass data are available only for $45 \leq A\leq 73$,
 whereas the theoretical mass data cover the range of $44 \leq A \leq 99$. 
Note that the theoretical mass data are not employed in Models 1a and  0a.
We can see unphysical jumps in the abundance at the boundary
 between  $A=73$ and 74  for Models 1a and 0a  in  Fig.~\ref{iso}.
The difference between Models 1a and 0a, on the other hand, arises from the different treatments of
 the bulk energies for the nuclei with the experimental mass data as well as of the internal degrees of freedom.
Note that the bulk energies including the temperature dependence given by Eq.~(\ref{eq:bulk})
 tend to be lower at non-vanishing temperatures,
 since hot nuclear matter has lower free energies than cold one due to excitations of nucleons. 
In other words, hot nuclei are more bound than cold ones
 owing to the increases in the internal degrees of freedom of the  nucleons inside nuclei.
The nuclei with  $45 \leq A\leq 73$ in Model~0a, for which the experimental mass data are available,
 have larger mass energies than the counter parts in Model~1a,
since the bulk energies of these nuclei in the former do not include the temperature dependence.
As a result, those nuclei are more abundant in Model~1a. 
On the other hand, the mass fractions of the nuclei with no experimental mass data ($A \geq 74$) in Model~0a 
 are larger than in Model~1a
 due to the double count of the temperature effect in the bulk energies $E_i^B(T)$ and internal degrees of freedom $g_i(T)$.
The difference between Models 0a  and 1a is clearer at higher temperatures, since the temperature dependences
in the bulk energies and internal degrees of freedom becomes stronger. 
We can also find that Model~2a do not produce the unphysical jumps in the abundances
 at both low and high temperatures owing to the modified treatment of the temperature effects 
as well as to  the employment of the theoretical mass data.

The average mass number of heavy nuclei $(Z\geq 6)$ as a function of density
 for the temperatures $(T=1, 5$ and 10~MeV) and proton fractions $(Y_p=0.3$ and $0.5)$
 is displayed in Fig.~\ref{ma} for Models 0a, 1a and 2a. 
It is found that for $T=1$~MeV the average mass number grows step-wise for Model~2a
even at high densities, $\rho_B \gtrsim 10^{13}$g/cm$^3$.
On the other hand, they grow monotonically in Models 0a and 1a at $A \gtrsim120$.
This is due to the lack of shell energies for the latter models. 
The nuclei in the vicinity of the neutron magic numbers ($N=28, 50, 82, 126, 184$) are abundant in Model~2a.
Note that the experimental mass data are available at the magic numbers $N=28,50,82$ under this condition,
whereas the mass data for the nuclei with $N=126$ such as $^{208}$Pb exist only near the stable line as shown in Fig.~\ref{nc}.
%This is the reason why we include more nuclear data with shell effects are included in the calculation thanks to KTUY theoretical mass data.
We can see that at the high temperatures ($T=5$, 10~MeV), 
Model~0a gives larger mass numbers than the other models.
This is because the nuclei, for which the mass data are available, are not abundant due to 
the lack of the temperature dependence in the bulk energies
and the heavier nuclei,
 for which the temperature dependence is taken into account but the shell effects are neglected, 
are abundant.
This feature can be confirmed also in the bottom panel of Fig.~\ref{iso}.

To summarize, the extended mass data and the temperature dependence 
remove the unphysical jumps found at the boundary between the nuclei with
available mass data and those without them, at high temperatures in our previous paper.
Even at low temperatures the wider use of the shell energy affects the nuclear abundances. 
%Fig.~? shows the iso abundances  1/
%%%%%%%%%%%%%%%%%%%%%%%%%%%%%%%%%%%%%%%%%%%%%%%%%%%%%%%%%%%%%
\subsection{Abundance of light nuclei} %KKK
In order to compare different models of light nuclei,
we employ the total mass fraction of deuteron, triton, helion and alpha particles, ${X_d+X_t+X_h+X_{\alpha}}$, for Models 2a, 2b and 2c.
The abundances of other light nuclei are not so large and will not be discussed in the following.
Fig.~\ref{lcmp}  shows the results for the temperatures $(T=1, 5$ and 10~MeV) and proton fractions $(Y_p=0.3$ and $0.5)$.
For $T=5$ and 10~MeV, we can see that the mass fraction of the light elements reaches the maximums at the densities $\rho_B \gtrsim 10^{12}$g/cm$^3$ in Model~2a.
This is because the light nuclei  in this model  have low bulk energies, since they are  calculated by the same LDM employed for heavy nuclei.
We can also see that  the light nuclei  are still abundant near the saturation densities for $T=10$~MeV 
due to the suppression of the surface energies of the light nuclei in the pasta phases as well as to the lack of the Pauli energy shifts.
Note that we assume in Models 2b and 2c that the light nuclei are quasi-particles  and do not form pastas.
The mass fractions of the light nuclei of Models 2b and 2c are similar between Models 2b and 2c
 because we do not adopt the LDM, which has a strong temperature dependence in the bulk energy.
The difference between Models 2b and 2c arises from the Pauli- and self-energy shifts.
We can see that the Pauli energy shifts slightly suppress the light nuclei at $T=$~5~MeV and $Y_p=0.5$ in Model~2b,
whereas the self-energy shifts make them more abundant  at $T=$~10~MeV in Model~2b  than  in Model~2c. 
Unlike in Model~2a, the light nuclei disappear near the saturation densities at $T=10$~MeV in these models,
since not only the self-energy shifts but also the Pauli energy shifts tend to suppress them
and free nucleons and heavy nuclei are dominant,  forming pastas. 
For $T=1$~MeV, the light nuclei dominate around $\rho_B \sim 10^{9}$g/cm$^3$. 
Since the Pauli- and self-energy shifts and the temperature dependence of bulk energies are 
rather minor, the three models give almost the same abundance.

\begin{figure}
\begin{center}
\begin{tabular}{ll}
\epsscale{.92}
   \plottwo{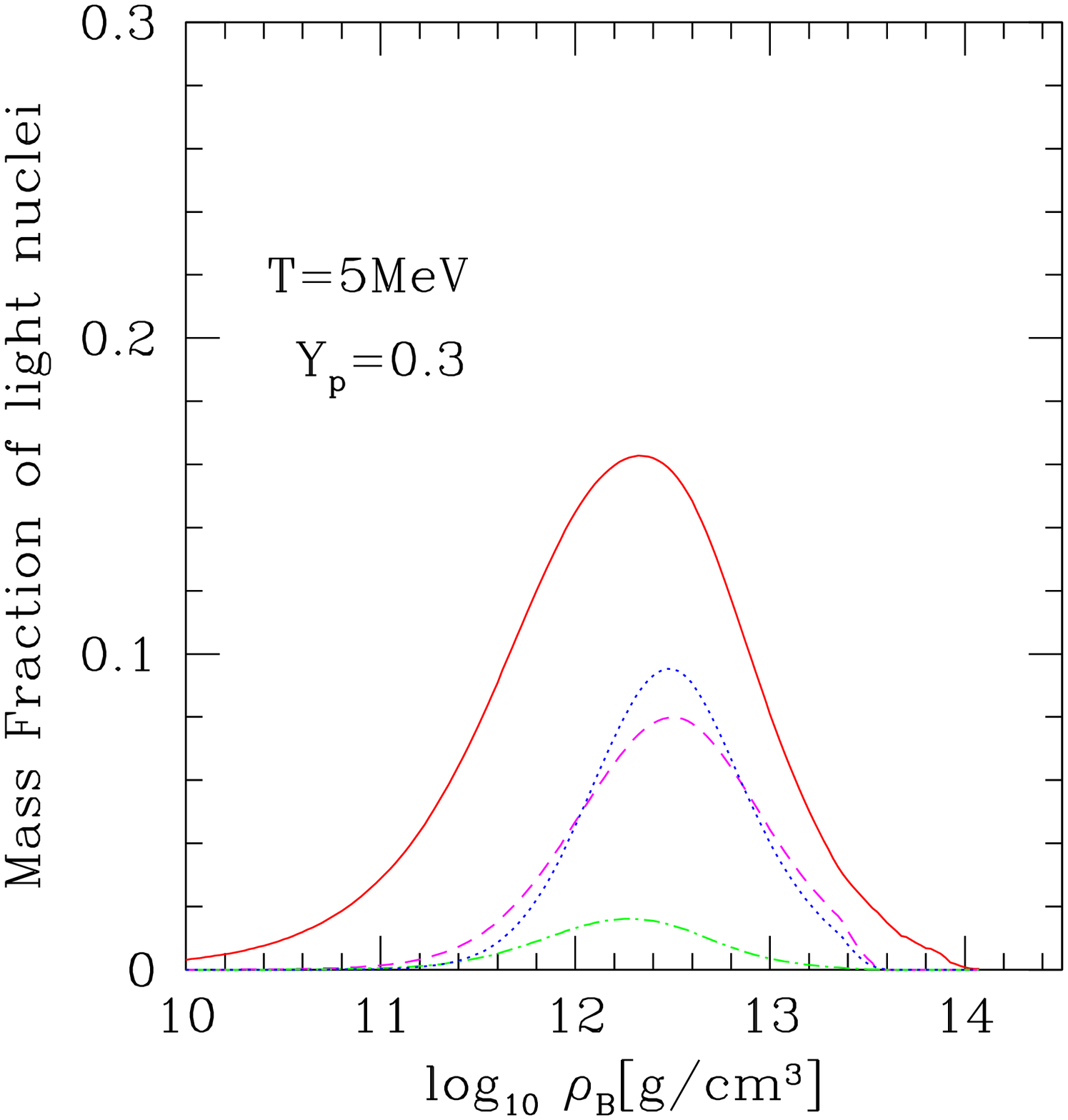}{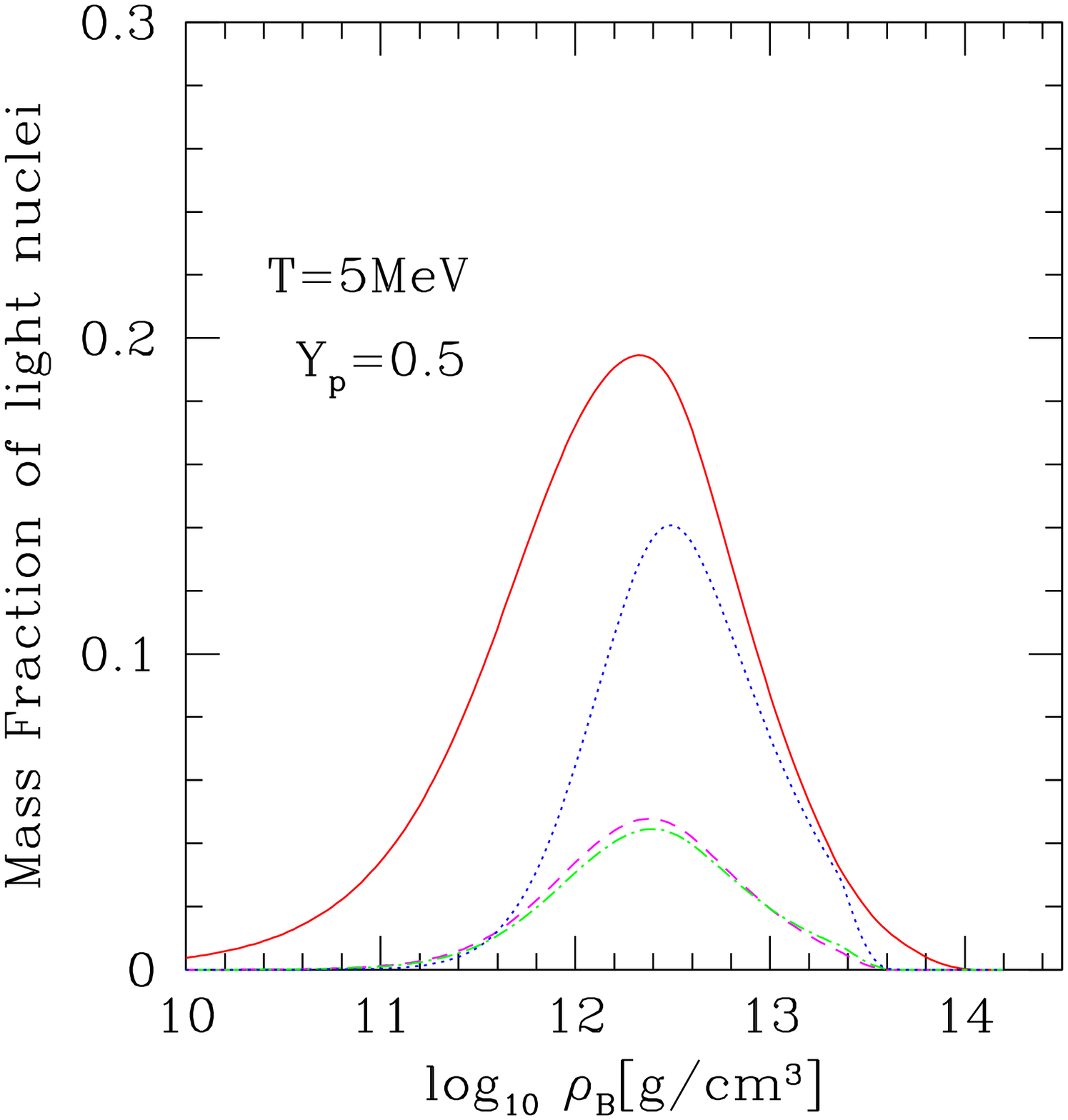} \\
   \plottwo{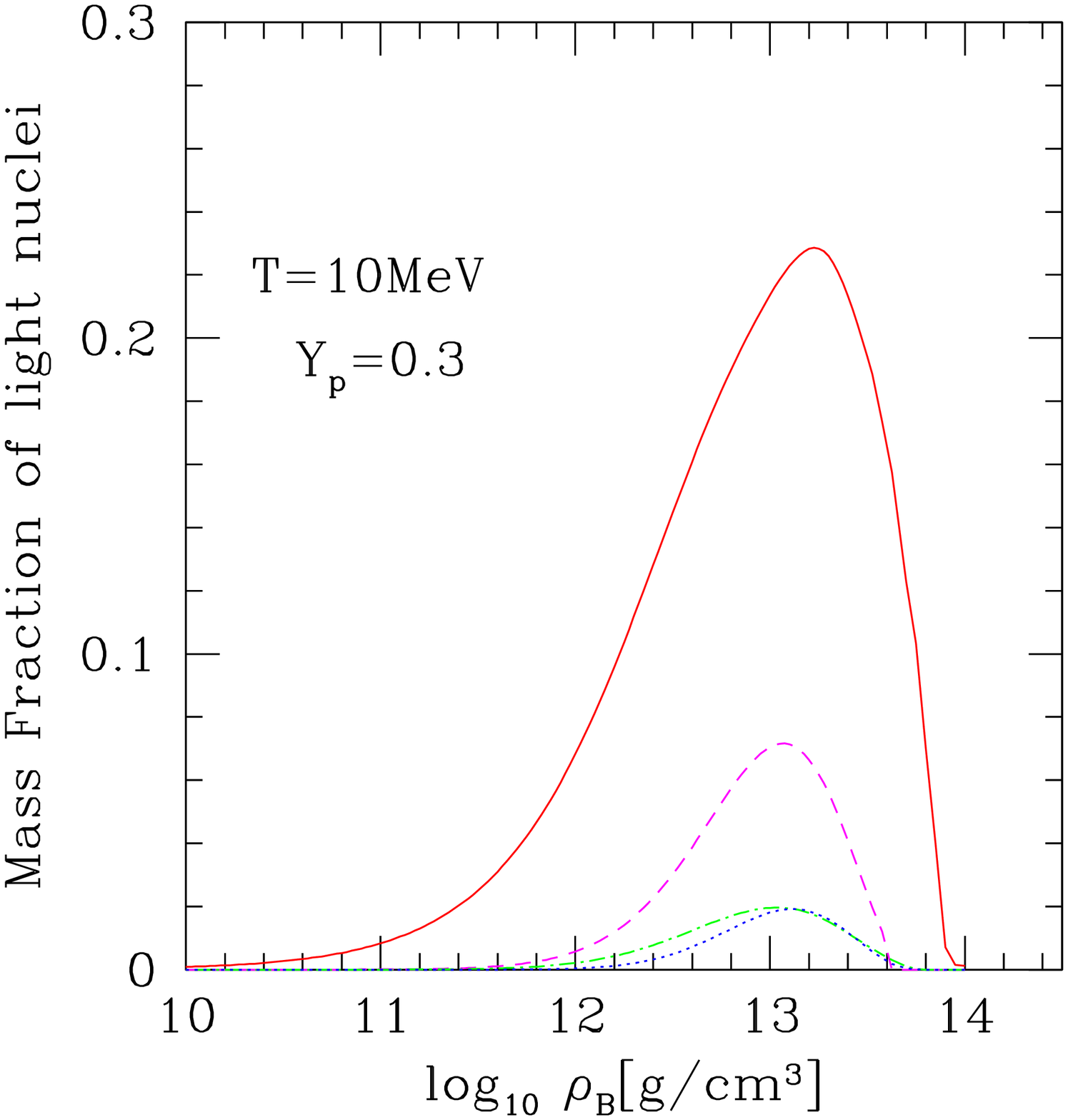}{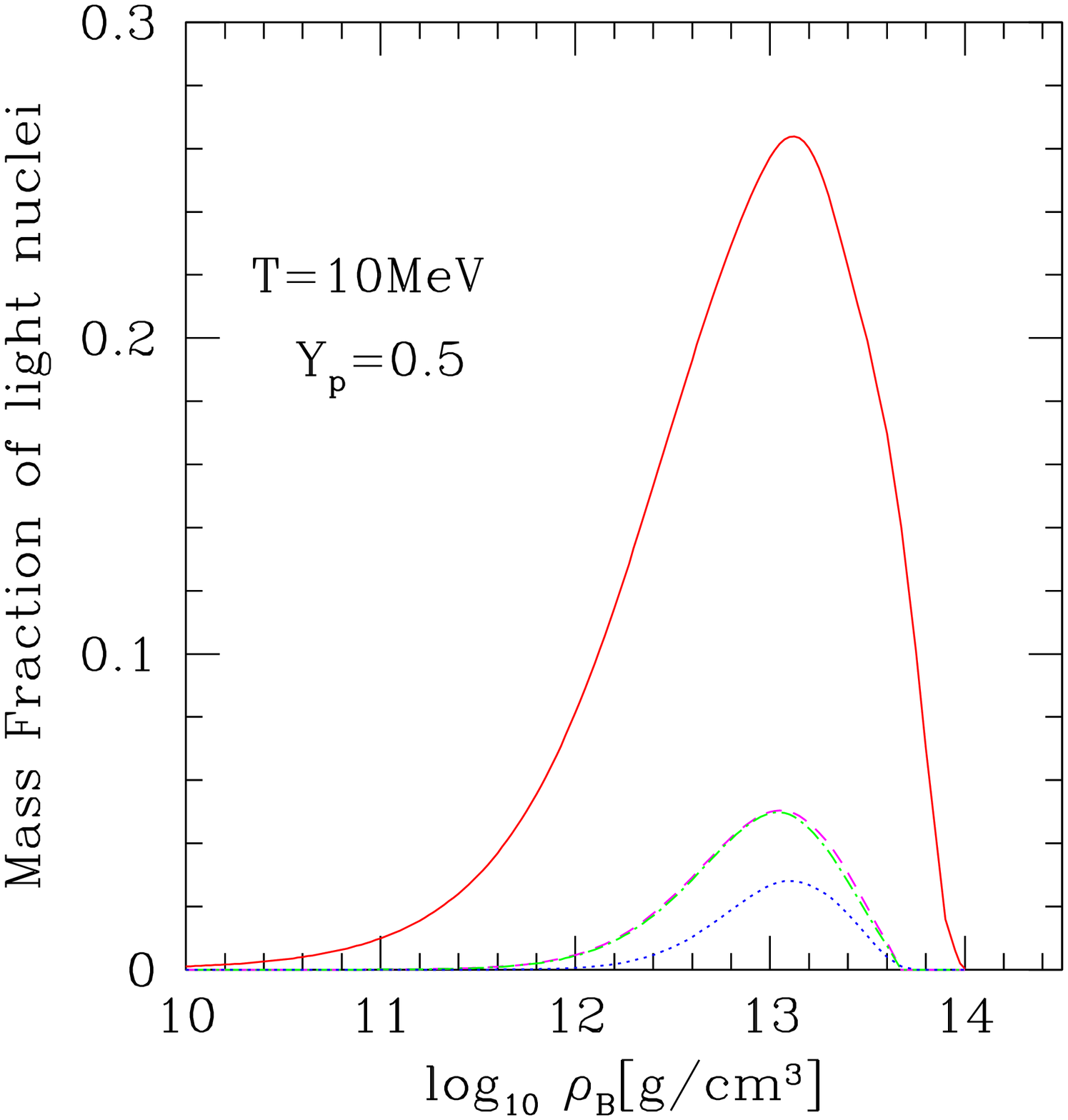} \\
    \end{tabular}
\end{center}
\caption{The mass fraction of each light nucleus for $d$ (red solid lines), $t$  (magenta dashed lines), $h$ (green dashed  dotted lines), $\alpha$ (blue doted lines) in Model~2b 
as a function of density for $T=5, 10 $~MeV and  $Y_p=0.3, 0.5$. }
\label{xl}
\end{figure}

We show the mass fraction of each light nucleus for Model~2b in Fig.~\ref{xl}.
For $T=10$~MeV, we can see that deuterons are the most abundant and alpha particles are  the least,
since lighter particles have more entropies per baryon. 
At $T=5$~MeV, deuterons still dominate.  
The alpha particles are also abundant, on the other hand,
 since the binding energy becomes also important in the minimization of the free energy density.
Note that alpha particles have the largest binding energy per baryon among the for light nuclei.
Under the neutron-rich condition of $Y_p=0.3$, the fraction of tritons is larger than that of helions,
whereas tritons and helions have almost the same abundance for the symmetric condition of $Y_p=0.5$.
At the lower temperature of $T=1$~MeV, though not shown in the figure, alpha particles are dominant  among the light nuclei 
due to the greatest binding energy per baryon.

We think that the abundance of light nuclei in Model~2a is too large at high temperatures due to the systematic overestimation
 of the binding energies in the LDM.
The Pauli- and self-energy shifts have influences on the light nuclei abundance at high temperatures ($T\gtrsim5$~MeV).
It is important that deuterons, tritons and helions can be as abundant as the alpha particles,
 which are normally assumed to be the representative light nucleus and incorporated  in the two standard  EOS's \citep{Lattimer1991,Shen1998,Hshen2011}.
%Fig.~? light nuclei 2/
%%%%%%%%%%%%%%%%%%%%%%%%%%%%%%%%%%%%%%%%%%%%%%%%%%%%%%%%%%%%%%%%%%%%%%%%%%%%%%%%%%%
\subsection{Thermodynamical quantities}
We compare the thermodynamics quantities for Models 0a,  2a and 2b.
Model~1a  is not presented because it is almost the same as Model~0a 
 at low temperatures and Model~2a at high temperatures ($T\geq $5~MeV). 
 
\begin{figure}
\begin{center}
\begin{tabular}{ll}
\epsscale{.82}
   \plottwo{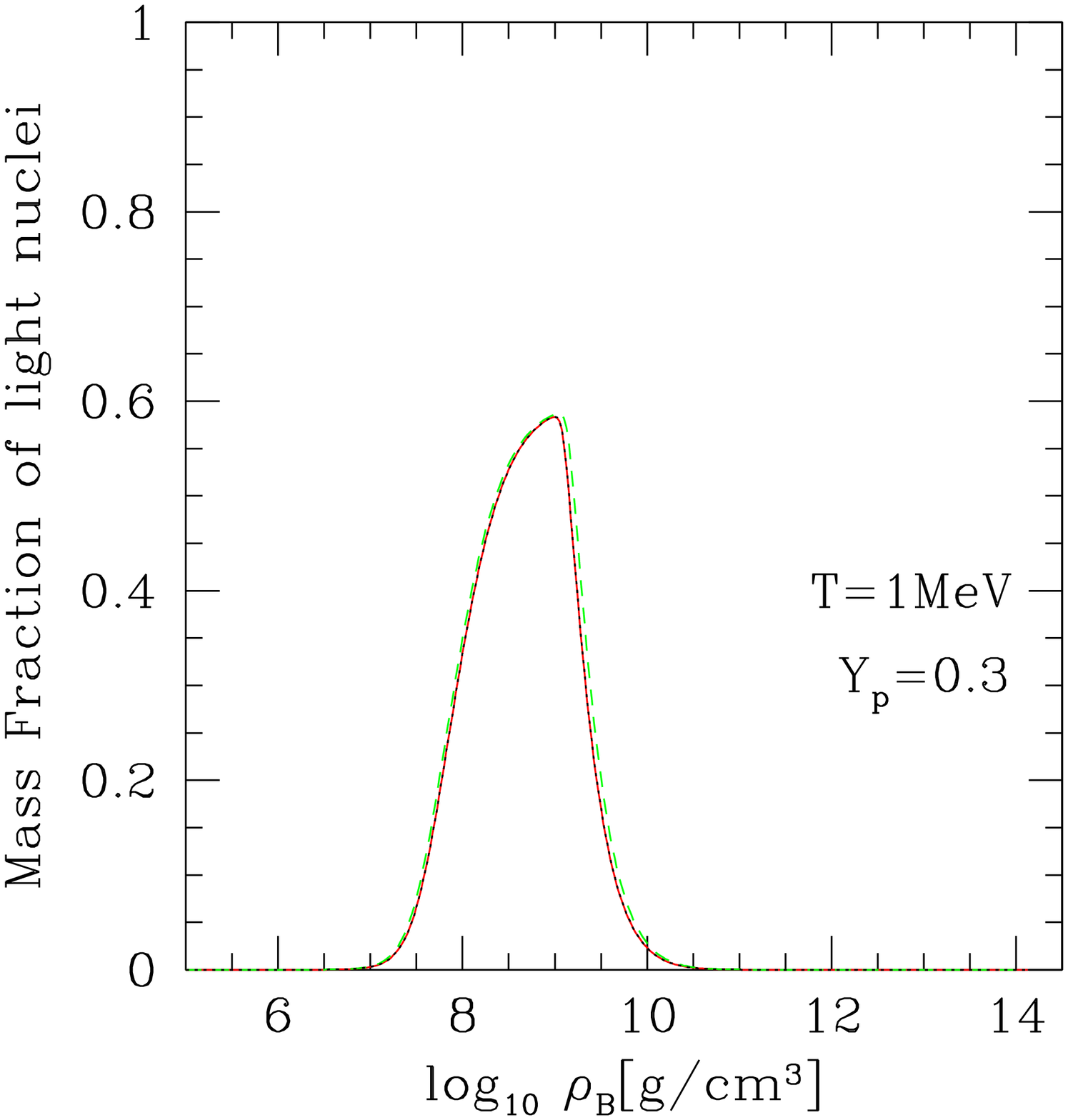}{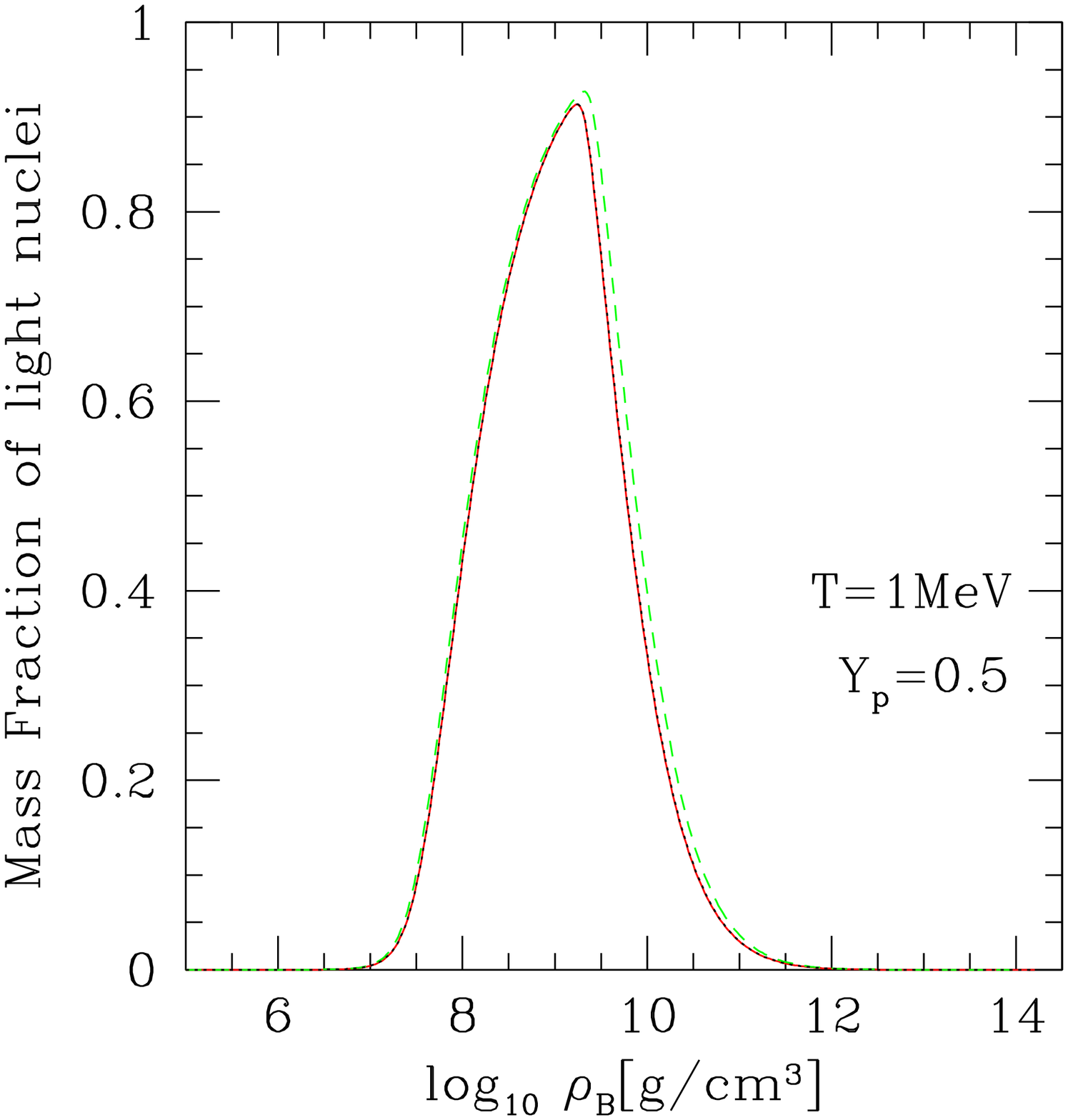} \\
   \plottwo{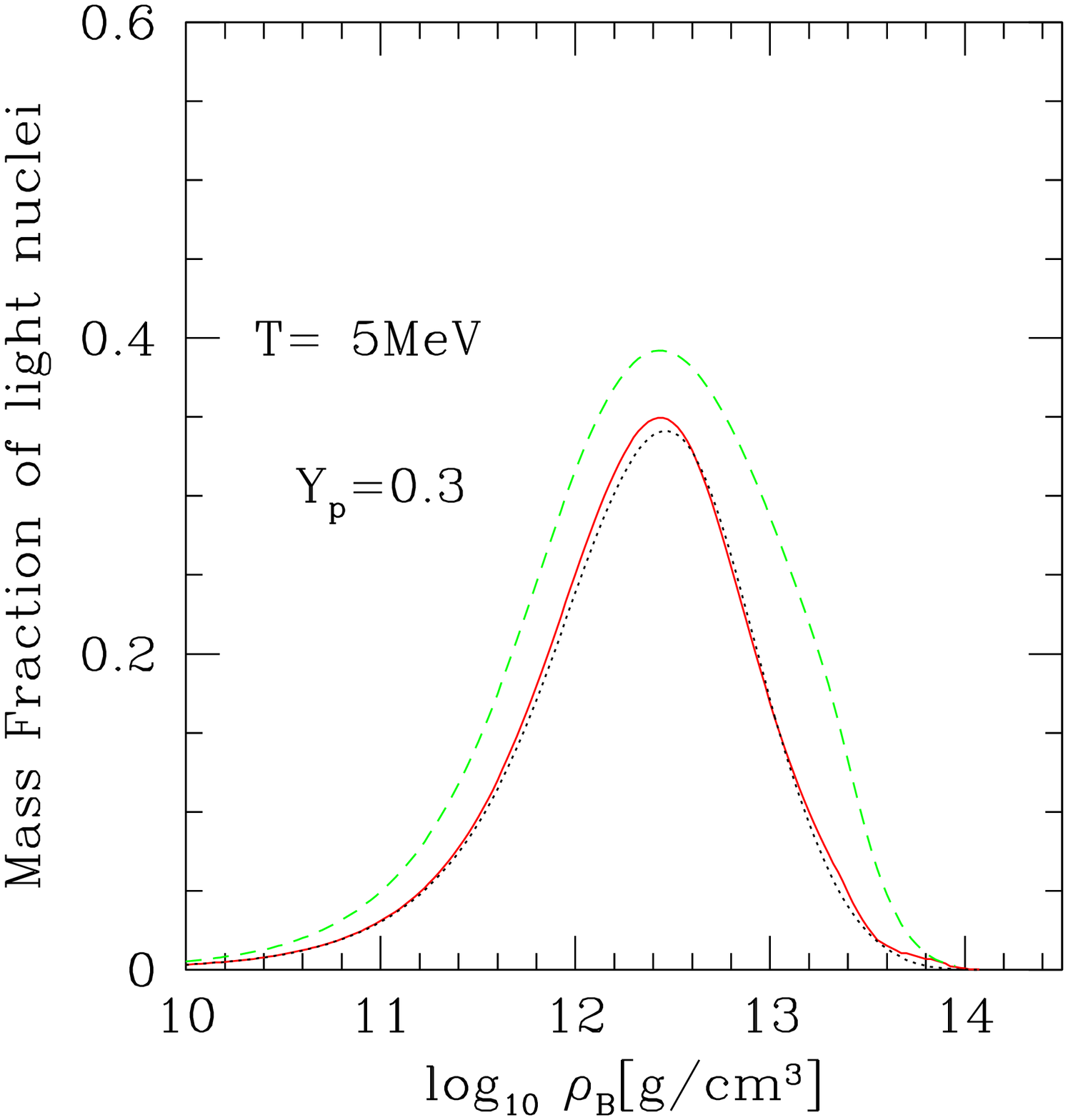}{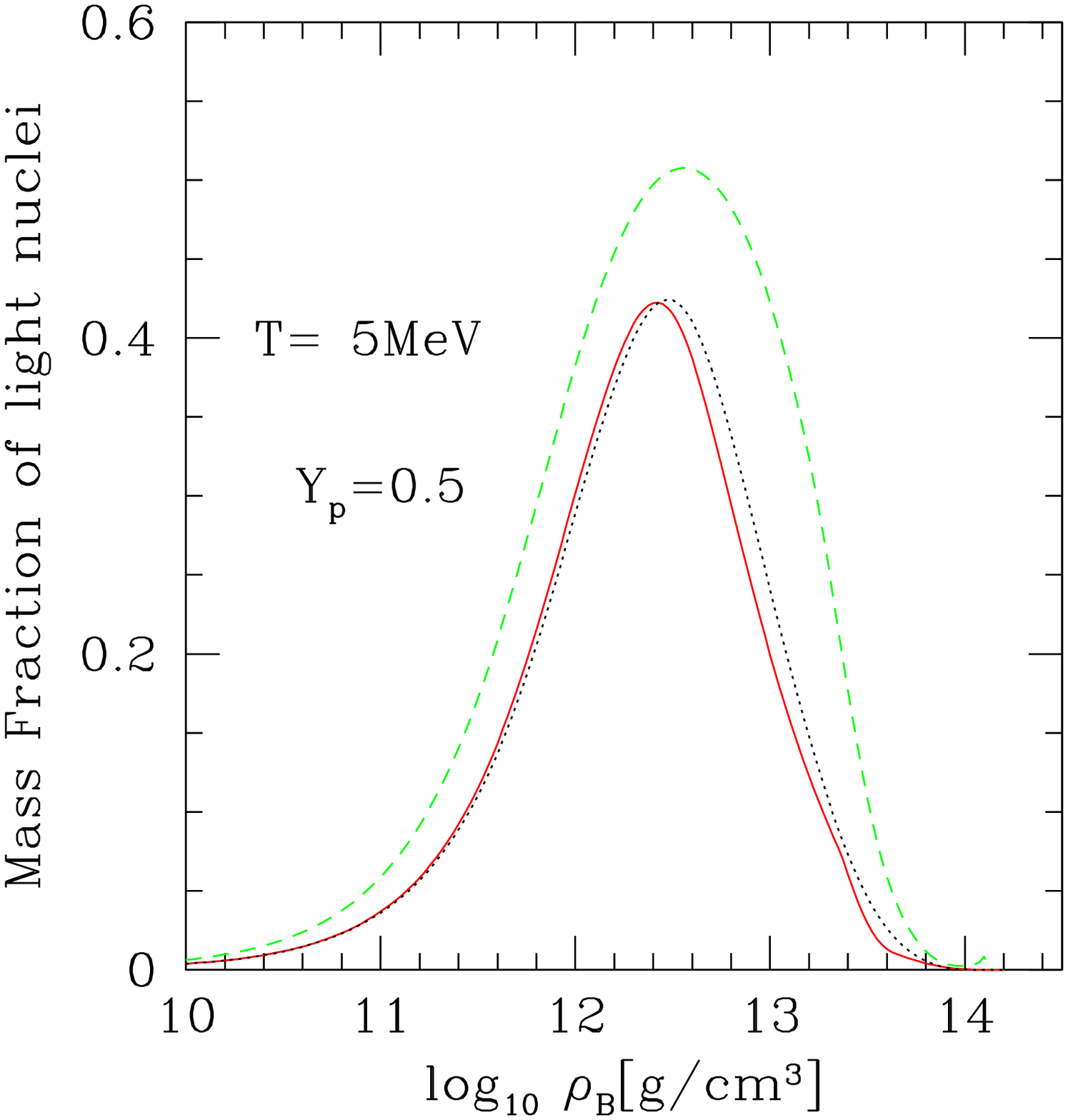} \\
   \plottwo{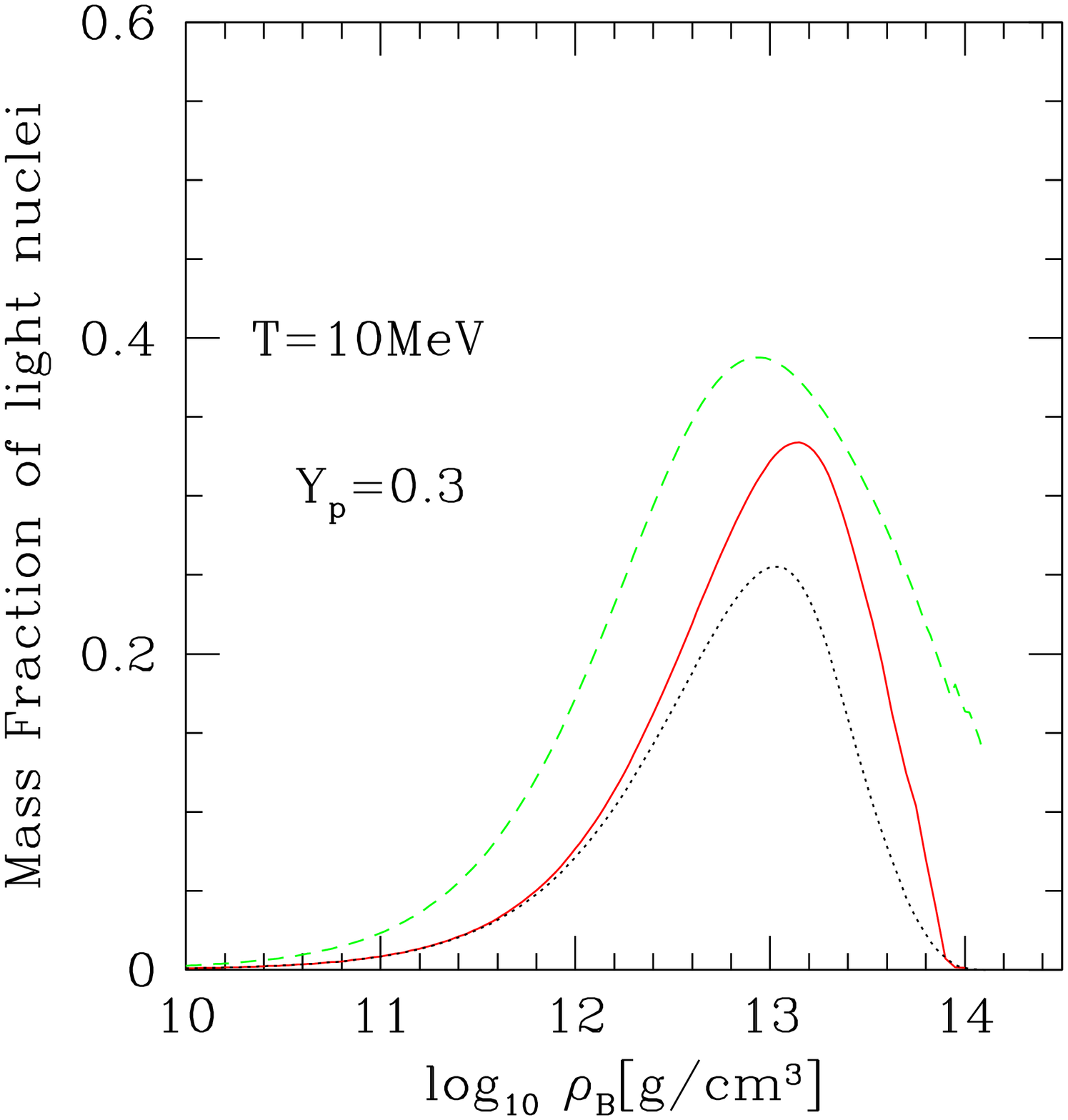}{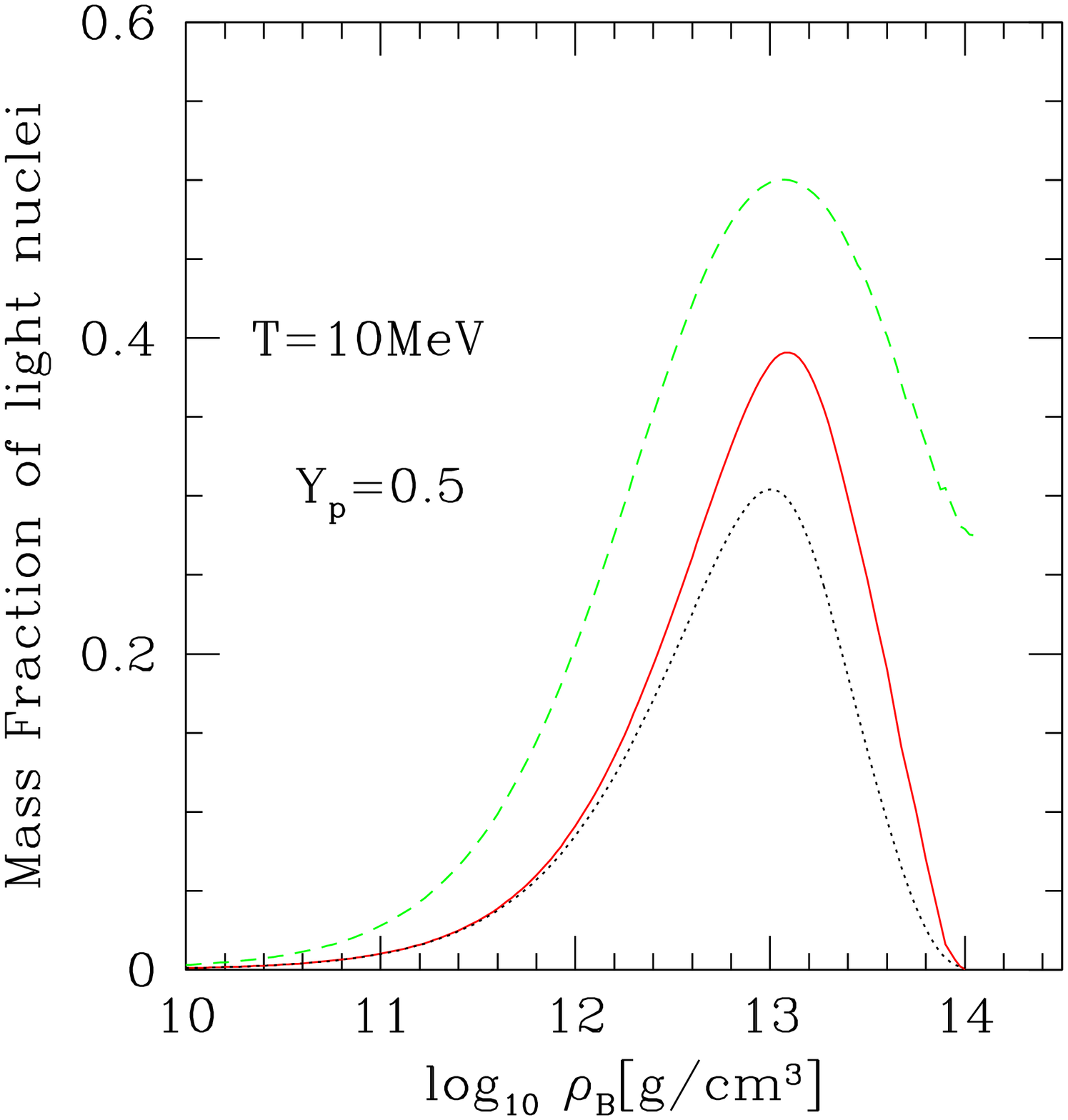} \\
    \end{tabular}
\end{center}
\caption{The mass fractions of light nuclei ($d, t, h, \alpha$) for Models 2a (green dashed lines), 2b (red solid lines) and 2c (black dotted lines) 
as a function of density for $T=1, 5, 10 $~MeV and  $Y_p=0.3, 0.5$. }
\label{lcmp}
\end{figure}

\begin{figure}
   \begin{center}
    \begin{tabular}{c}
               \resizebox{63mm}{!}{\plotone{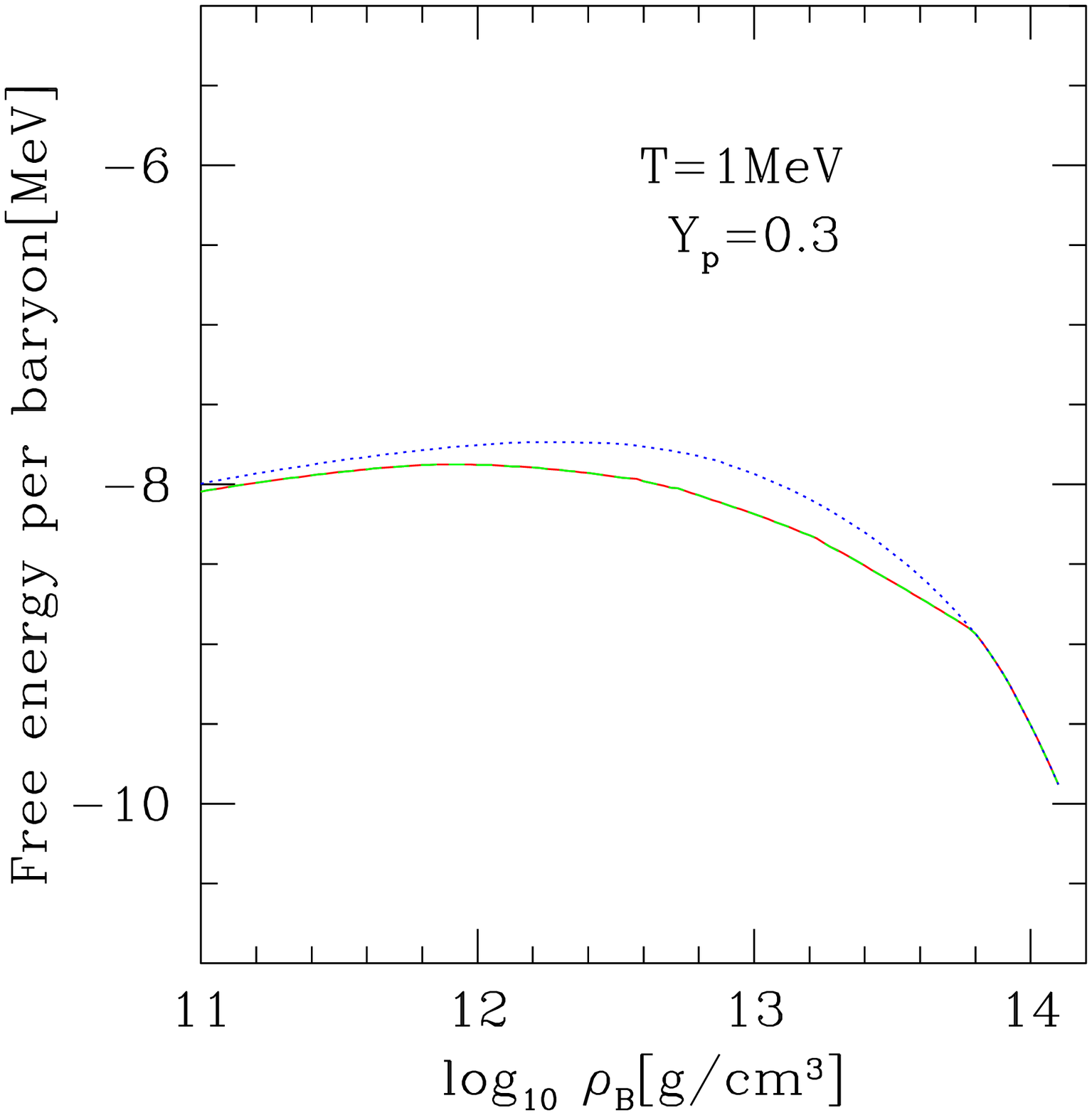}} \\
               \resizebox{63mm}{!}{\plotone{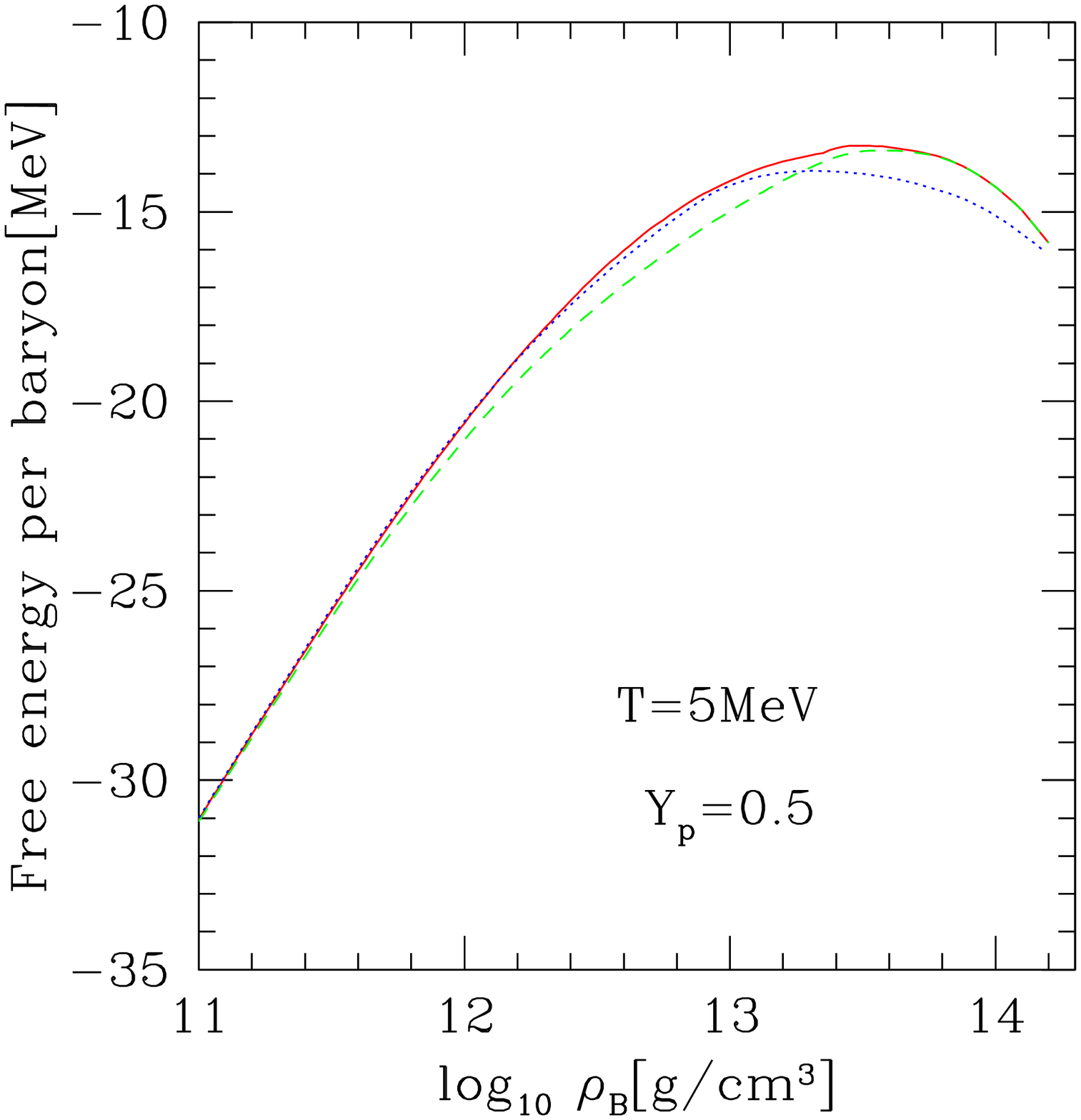}} \\
               \resizebox{63mm}{!}{\plotone{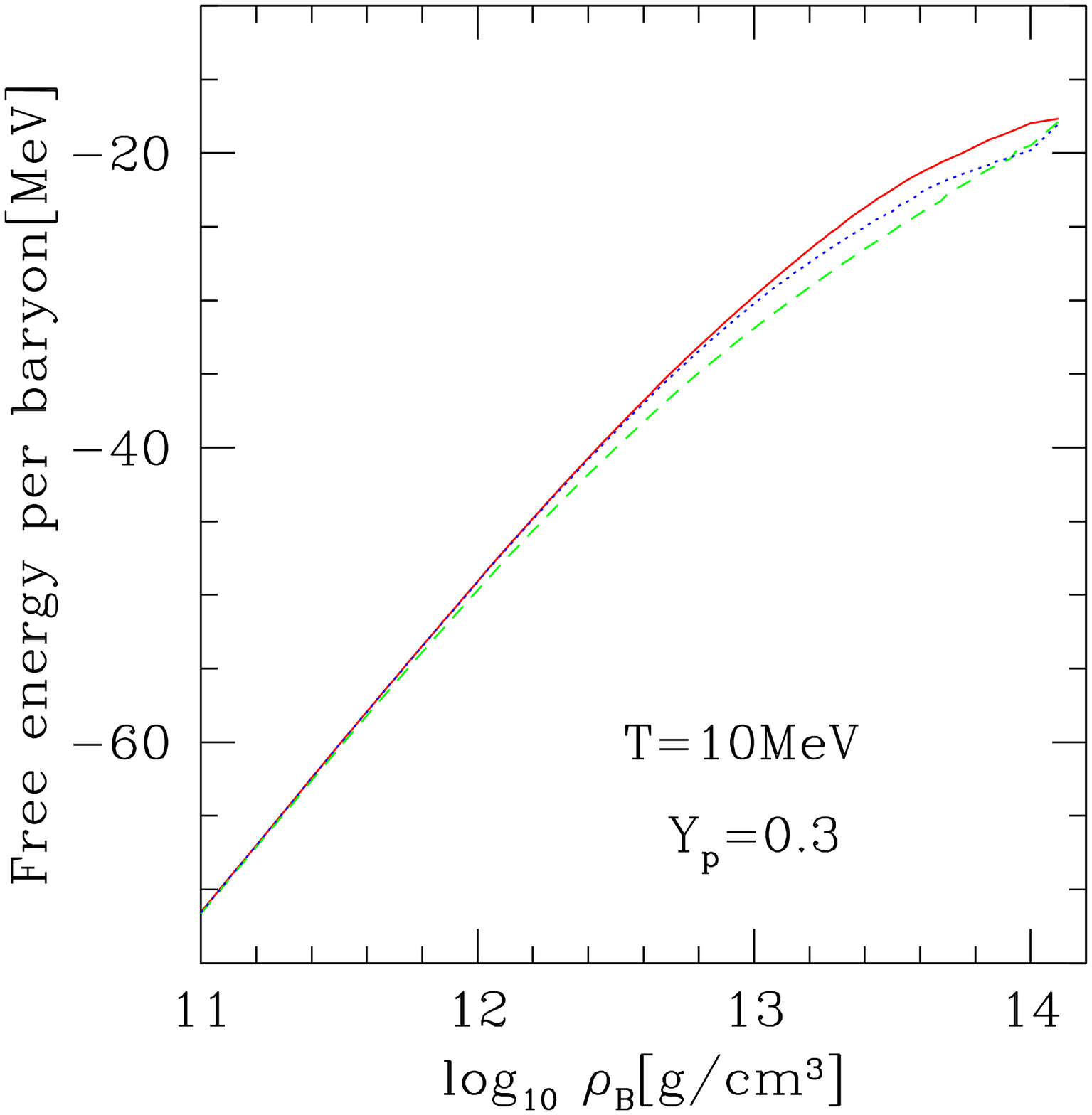}} \\
    \end{tabular}	
   \end{center}
\caption{The free energy per baryon for  Models 0a  (blue dotted lines), 2a (green dashed lines), 2b (red solid lines)  as a function of 
density for ($T=1$~MeV, $Y_p=0.3$), ($T=5$~MeV, $Y_p=0.5$) and ($T=10$~MeV, $Y_p=0.3$) from top to bottom.}
\label{fr}
\end{figure}

Fig.~\ref{fr} shows the free energies per baryon as a function of density for the three combinations of temperature and proton fraction:
($T=1$MeV, $Y_p=0.3$), ($T=5$MeV, $Y_p=0.5$), ($T=10$~MeV, $Y_p=0.3$). 
For $T=1$~MeV, Model~0a has the highest free energy
due to the  lack of the theoretical mass data, which are also evident  in Figs.~\ref{dis0},~\ref{iso}~and~\ref{ma}.
Model~0a  neglects the shell energies of the nuclei, for which no experimental mass data are available, 
 whereas Models 2a and 2b employ the theoretical mass data.
The free energies per baryon are not so different at $T=1$~MeV
 between Models 2a and 2b, because the binding energies of heavy nuclei are dominant at this low temperature.
For $T=5$ and 10~MeV, on the other hand, we find that Model~2a gives lower free energies than Model~2b.
The difference originates from the fact that the light nuclei are more abundant in Model~2a than in Model~2b,
since  Model~2a gives lower bulk energies  to light nuclei due to the strong temperature dependence as shown in Fig.~\ref{lcmp}.
Model~0a  also gives lower free energies  per baryon than Model~2b 
because of the double count of the temperature effects in the bulk energies $E^B_i(T)$ and internal degrees of freedom $g_i(T)$ of the nuclei,
 for which no experimental data exist.

\begin{figure}
   \begin{center}
    \begin{tabular}{c}
               \resizebox{63mm}{!}{\plotone{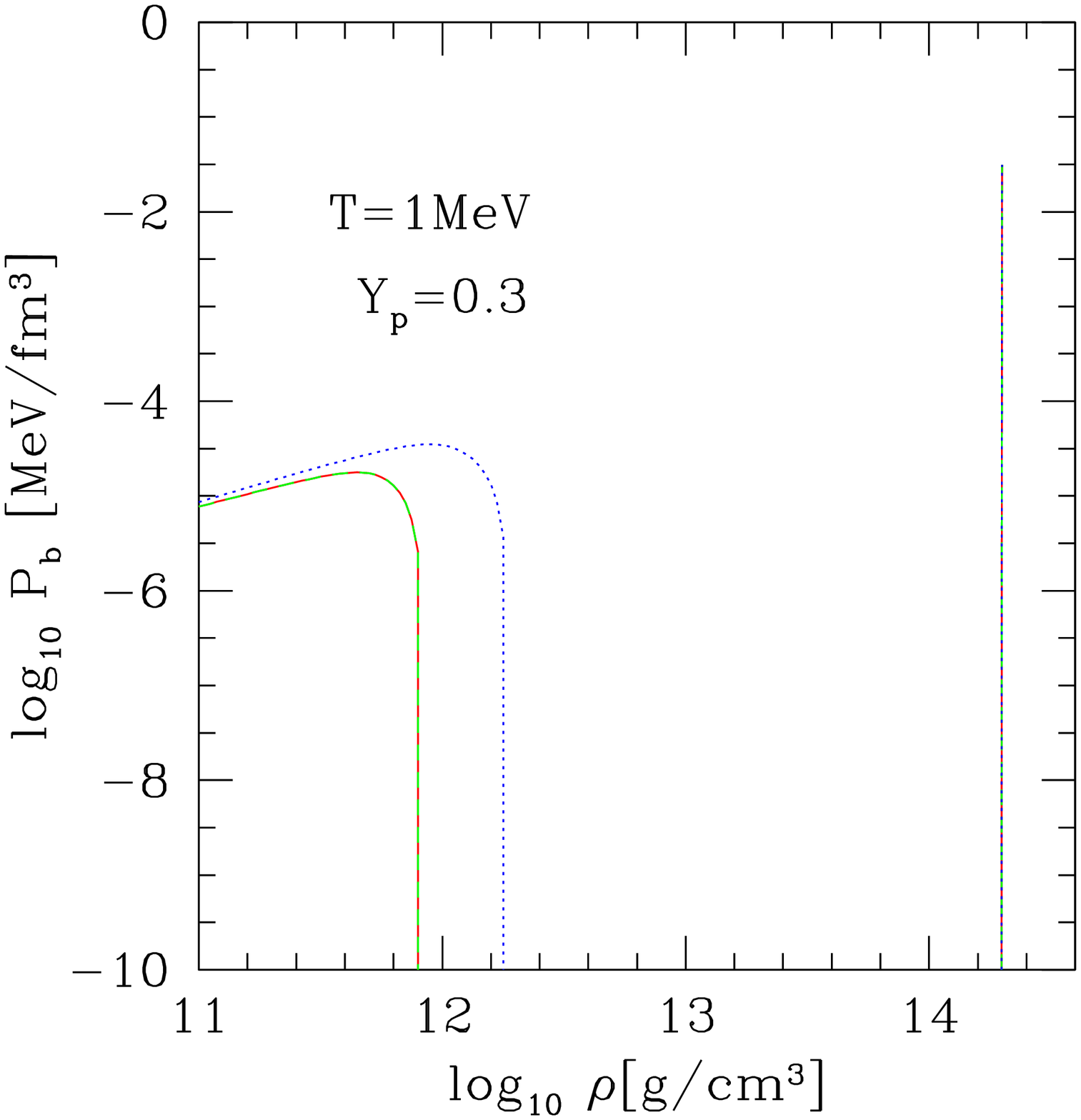}} \\
               \resizebox{63mm}{!}{\plotone{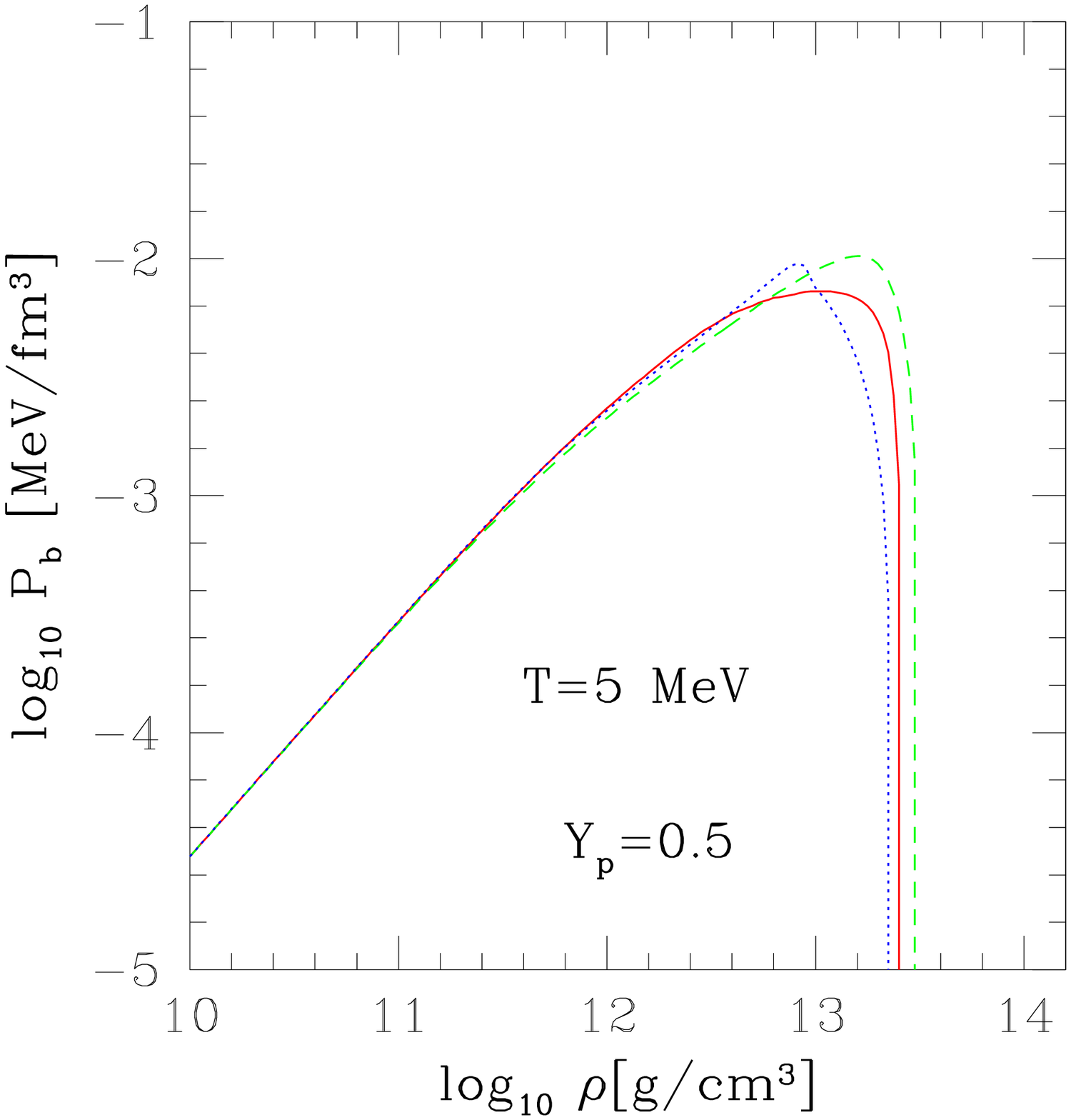}} \\
               \resizebox{63mm}{!}{\plotone{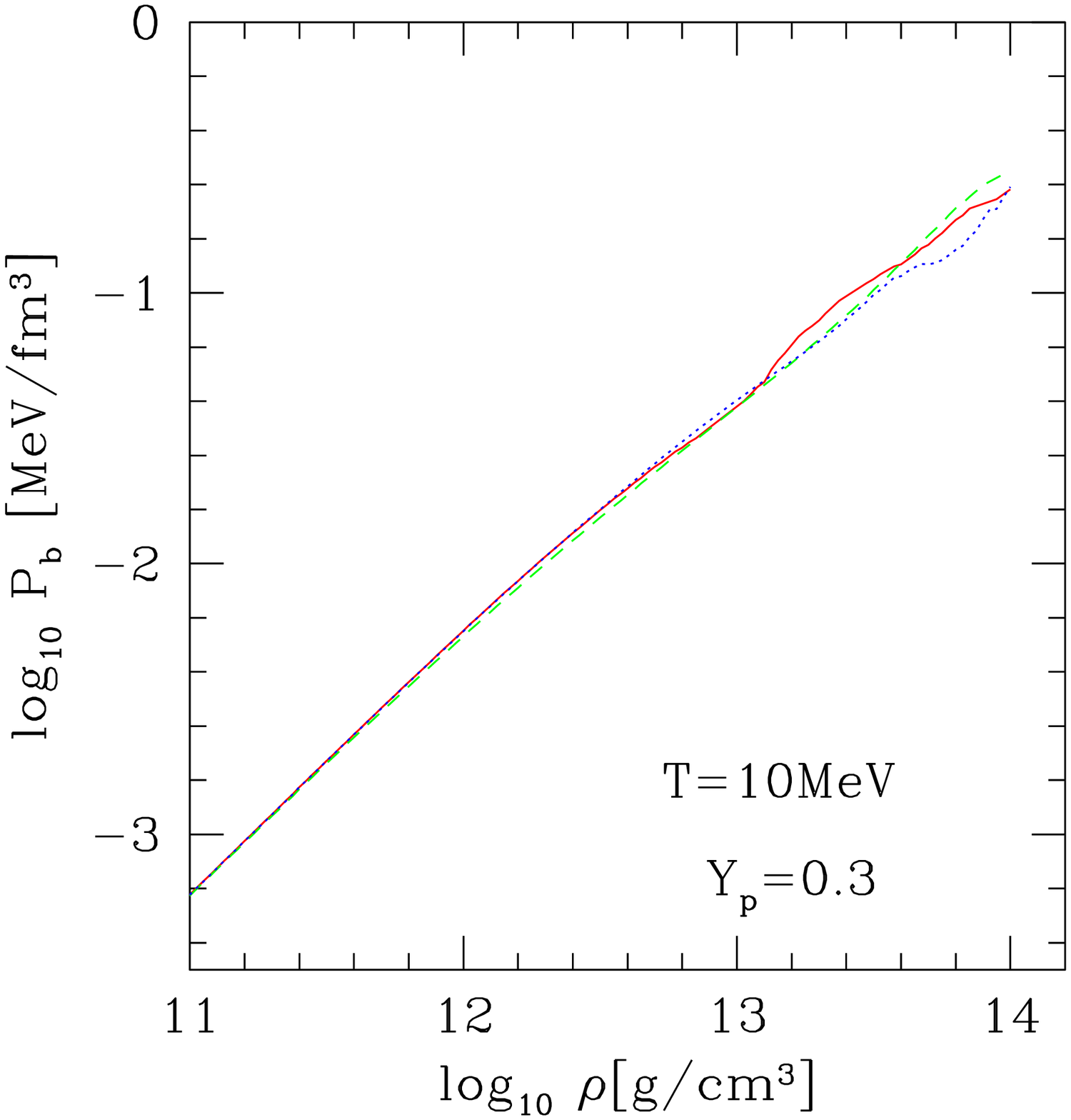}} \\
    \end{tabular}	
   \end{center}
\caption{The baryonic pressure for  Models 0a (blue dotted lines), 2a (green dashed lines), 2b (red solid lines) as a function of 
density for ($T=1$~MeV, $Y_p=0.3$), ($T=5$~MeV, $Y_p=0.5$) and ($T=10$~MeV, $Y_p=0.3$) from top to bottom.}
\label{pr}
\end{figure}

The pressure is shown as a function of density for three combinations of temperature and proton fraction,
($T=1$~MeV, $Y_p=0.3$), ($T=5$~MeV, $Y_p=0.5$), ($T=10$~MeV, $Y_p=0.3$), in Fig.~\ref{pr}. 
The three models agree with one another at low densities, $\rho_B \lesssim 10^{12}$g/cm$^3$. 
For $T=1$ and 5~MeV, the baryonic pressure is negative when the Coulomb-energy contribution, 
which is negative owing to the attractive Coulomb interactions between protons inside nuclei and uniformly-distributed electrons (the so-called Coulomb corrections),
dominates over the other positive contributions. 
For $T=1$~MeV, the density, at which the pressure drop occurs in Model~0a,  is higher than in the other models.
The average mass numbers are smaller in Model~0a 
 as shown in Fig.~\ref{ma} due to the lack of the theoretical mass data
and the pressure drop occurs at higher densities than the other models. 
The pressures   are almost the same at $T=1$~MeV between Models 2a and 2b,
since the contribution of light nuclei is negligible at low temperatures.  
For $T=5$~MeV, on the other hand,
the density, at which the pressure drop occurs, is the lowest in Model~0a.
This is because the average mass numbers  are larger as shown in Fig.~\ref{ma}.
The density, at which the of pressure drop occurs, is highest in Model~2b, since the abundance of light nuclei,
 which contributes positively to the pressure, is larger in Model~2a than in Models 0a  and 2b as shown in Fig.~\ref{lcmp}.
At the even higher temperature  of $10$~MeV, the positive thermal pressures of free nucleons and nuclei are dominant.
Model~2a gives a little higher pressure than Model~2b near the saturation density, since the light nuclei are most abundant in Model~2a as shown in Fig.~\ref{lcmp}.
For Model~2b the pressure is the highest at $ \rho_B \sim 10^{13}$g/cm$^3$,
 because free nucleons are the most abundant among the three models.

\begin{figure}
   \begin{center}
    \begin{tabular}{c}
               \resizebox{63mm}{!}{\plotone{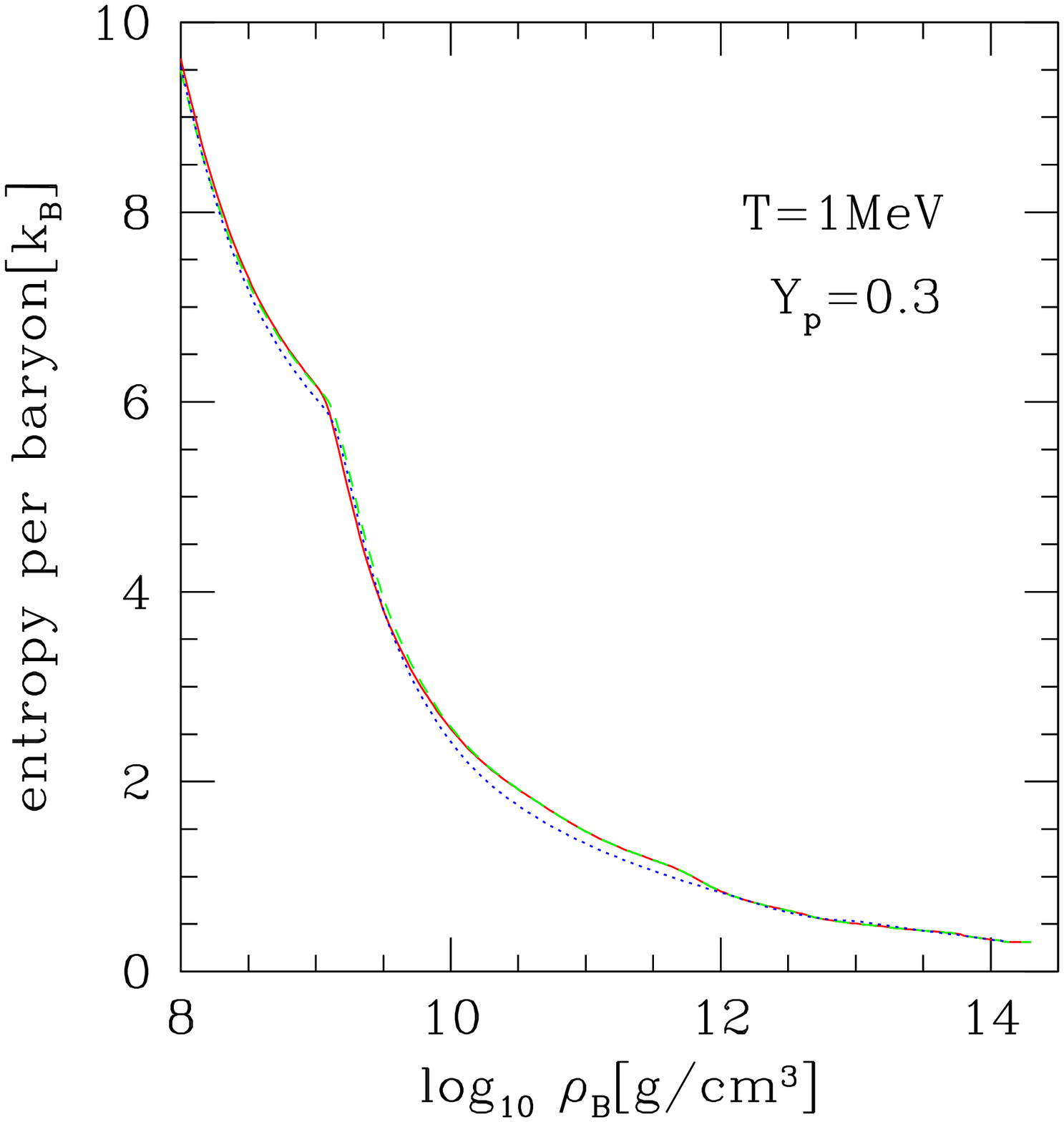}} \\
               \resizebox{63mm}{!}{\plotone{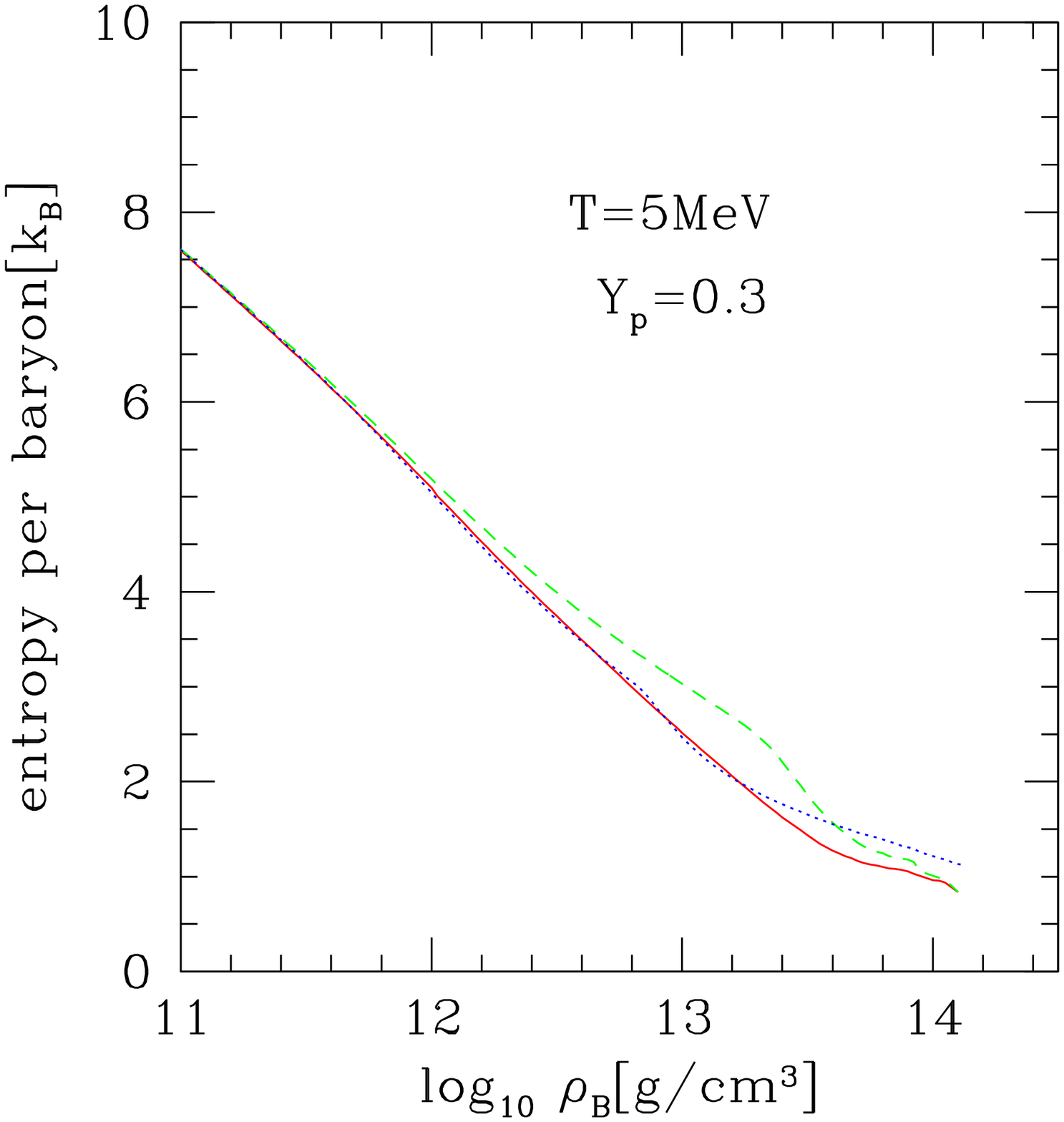}} \\
               \resizebox{63mm}{!}{\plotone{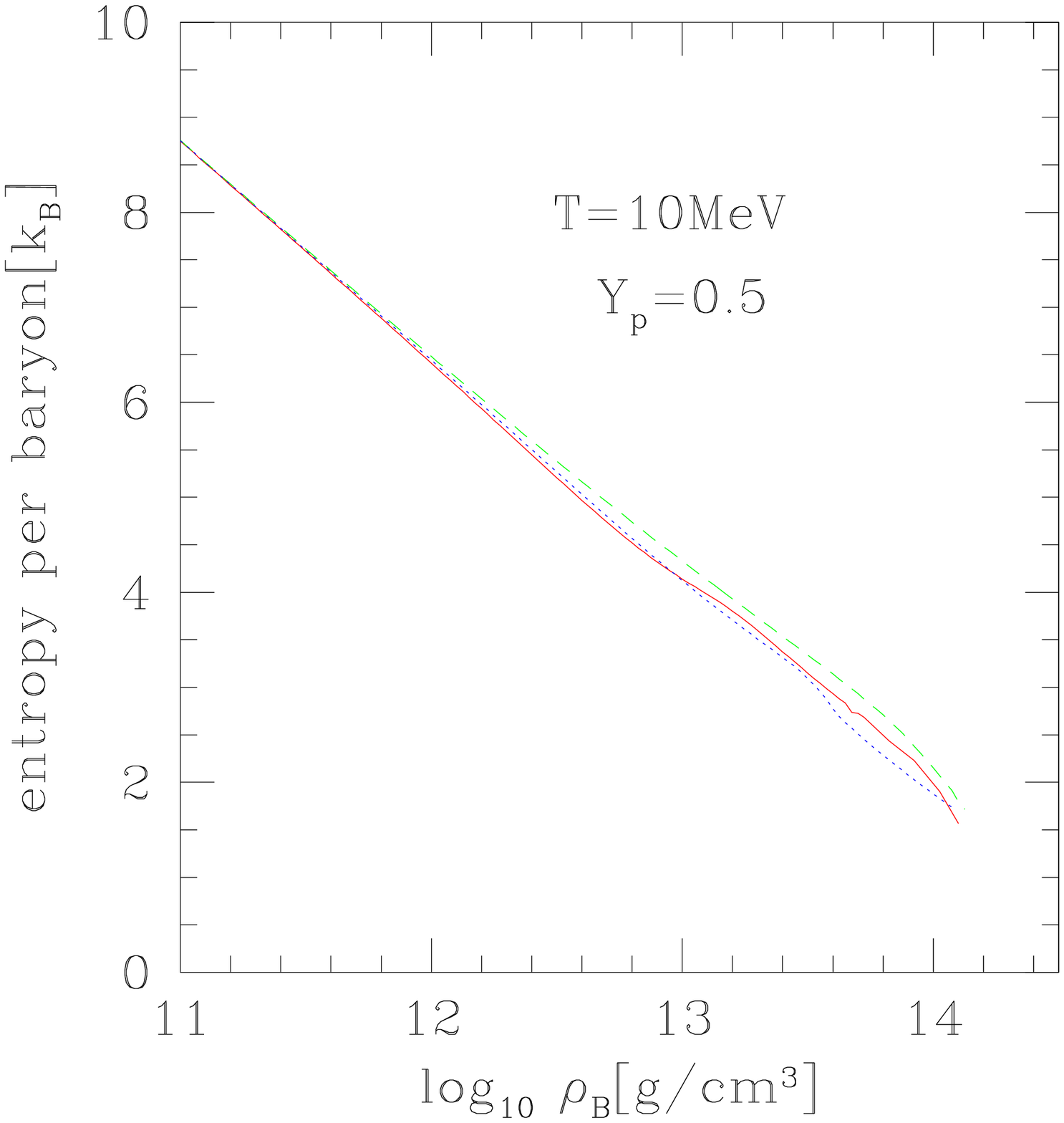}} \\
    \end{tabular}	
   \end{center}
\caption{The entropy per baryon for Models 0a  (blue dotted lines), 2a (green dashed lines), 2b (red solid lines) as a function of 
density for ($T=1$~MeV, $Y_p=0.3$), ($T=5$~MeV, $Y_p=0.3$) and ($T=10$~MeV, $Y_p=0.5$) from top to bottom.}
\label{en}
\end{figure}

The entropy per baryon is displayed as a function of density for three combinations of temperature and proton fraction,
($T=1$~MeV, $Y_p=0.3$), ($T=5$~MeV, $Y_p=0.3$) and ($T=10$~MeV, $Y_p=0.5$), in Fig.~\ref{en}. 
For $T=1$~MeV, the entropy per baryon is almost identical among the three models.
For $T=5, 10$~MeV, on the other hand, Model~2a has larger values than Models 0a  and 2b at $\rho_B \sim 10^{12}$g/cm$^3$ owing to
the larger population of light nuclei as shown in Fig.~\ref{lcmp}.
For $T=5$~MeV, Model~0a has the highest entropy per baryon near the saturation density  because of the double count of temperature effects for the heavy nuclei with no mass data.

\subsection{Phase diagram}
\begin{figure}
   \begin{center}
    \begin{tabular}{c}
         \includegraphics[width=95mm]{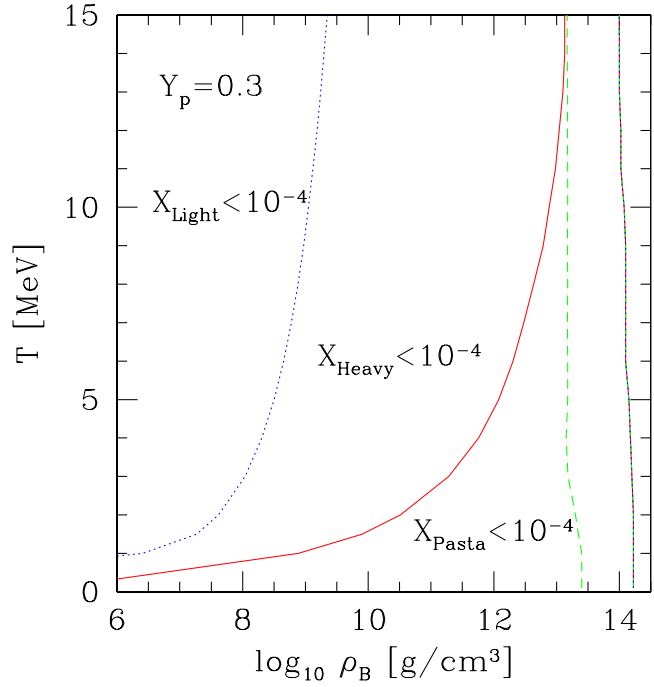}
    \end{tabular}
\caption{ Phase diagram of Model~2b at $Y_p = 0.3$ in the $\rho_B, T$ plane.   
Blue dot lines show the boundary where the light nuclei fraction ($Z\leq 5$) $X_L$ changes between $X_L <10^{-4} $ and $X_L > 10 ^{-4}$.
Red solid lines are that of heavy nuclei $(Z\geq 6)$. Green dashed  lines are that of pasta phase nuclei $X_{Pasta} =\Sigma X_{i}$ for pasta nuclei $(u_i>0.3)$}
    \label{pd}		
   \end{center}
\end{figure}

We finally discuss the phase diagram for Model~2b, which indicates the region where each of light, heavy and pasta nuclei is abundant 
 boundary for the change of dominant composition.
 The boundaries are chosen so that each fraction of light, heavy and pasta nuclei would be $10^{-4}$ following  \citet{Shen1998}.
The total mass fraction of heavy nuclei is evaluated as $X_H=\sum_{Z_i\geq 6} X_i$  and
 that of light nuclei is $X_L=\sum_{Z_j\leq5} X_j$, where $X$ means the mass fraction.                
The mass fraction of the pasta nuclei $X_{Pasta}$ is also calculated as
\begin{equation}
\label{pasta}
  X_{Pasta}  = \sum_{u_i>0.3} X_ i,
\end{equation}
where  $u_i$ is the volume fraction of nucleus $i$ in its Wigner-Seitz cell.
Note that $d, t, h,$ and $\alpha$ are not included in this summation, since they are assumed not to form the pastas.
In our EOS, the  nuclei  with  $u_i\leq 0.3$ are assumed to be normal,
 whereas those with  $0.7\leq u_i<1.0$ are supposed to be bubbles  and $u_i=1.0$  corresponds to the uniform matter. 
The nuclei with $0.3<u_i<0.7$  are interpolated between the droplets and bubbles,
 a very crude approximation to the rod, slab anti-rod phases.
We can see in Fig.~\ref{pd} the density range, in which heavy nuclei are abundant, becomes narrower as the temperature rises.  
This is because the entropy term  $-TS$ of free nucleons and light nuclei become more important than the internal energy term $U$ of heavy nuclei in the free energy $F=U-TS$.
Near the saturation densities,  however, the pasta phase survives even at high temperatures,
since it has almost the same free energies per nucleon as  uniform matter.
The abundance of pasta nuclei decreases and that of free nucleons increases monotonically
as the temperature get higher due to the small surface and Coulomb energies of the pasta phase.

\section{Summary and Discussions}
We have extended the baryonic equation of state at sub-nuclear densities,
 which was developed for the use in core-collapse supernova simulations in our previous paper.
The EOS provides the abundance of various nuclei up to the proton number of $1000$ in addition to thermodynamical quantities.
The major modifications in  the new EOS include the different treatments 
of the bulk and shell energies of heavy nuclei and the internal degrees of freedom,
 the use of the theoretical mass data wherever available, 
and the adoption of the different estimation of the masses of the light nuclei based on the quantum approach.
The bulk energies of all heavy nuclei $(Z\geq6)$ now have the temperature dependence,
 which is different from the previous one. 
As a matter of fact, 
the temperature effects are encapsulated only in the internal degree of freedom of the nuclei,
for which mass data are available, in the previous paper.
In this paper, we employ the theoretical mass data in addition to the experimental ones to obtain the shell energies.
For the light nuclei with $Z \leq$ 2 and $N \leq$ 2, the results of quantum
calculations are adopted to better reproduce the binding energies of those nuclei at high densities and temperatures.
For other light nuclei ($Z \leq 5$), we use the mass formula based on the LDM,
which is different from the one for heavy nuclei.
The LDM for the light nuclei gives a temperature dependence of the binding energies 
similar to that obtained from the quantum approach for $Z \leq$ 2 and $N \leq$ 2.
 %KTUY

The basic part of the the model free energy density is the same as that given in \citet{Furusawa2011}.  
This model free energy density is constructed so that it should reproduce the ordinary NSE results at low densities
 and make a continuous transition to the supra-nuclear density EOS obtained from the RMF. 
For the nuclei with neither experimental nor theoretical mass data available, 
we have neglected the shell energies. 
At high densities, where the nuclear structure is affected
 by the presence of other nuclei, nucleons and electrons, 
we have reduced by hand the shell energy from the value obtained from the experimental or theoretical data to zero at high densities. 
Assuming the charge neutrality in the W-S cell, we have calculated the Coulomb energy of nuclei. 
Close to the nuclear saturation density,
 the existence of the pasta phase has been taken into account
in calculating the surface and Coulomb energies. 
The free energy density of the nucleon vapor outside nuclei is calculated by  the RMF employed for the description of heavy nuclei. 

For some representative combinations of density, temperature and proton fraction,
 we have made a comparison of the abundances of nuclei 
as well as thermodynamical quantities obtained in different models. 
%There are two major differences between the new and old EOS.
The model without the temperature dependence in the bulk energies for the nuclei
 with experimental mass data available (Model~0a) 
yields the unphysical jumps in isotope distributions, especially at high temperatures,
because the bulk energies obtained from the RMF theory are lower than the experimental values. %rp18
We have found that the introduction of the theoretical mass data solves this problem and changes
the mass fractions as well as the average mass numbers.
We have also revealed that the new EOS including the Pauli- and self-energy shifts
give lower abundances of the light nuclei  than the old EOS based on the LDM.
This is because the LDM overestimates the binding energies of the light nuclei at high temperatures.
The Pauli  and self-energy shifts also affect the light nuclei abundance at high temperatures and densities.

\begin{figure}
\begin{center}
\begin{tabular}{ll}
\epsscale{.82}
   \plottwo{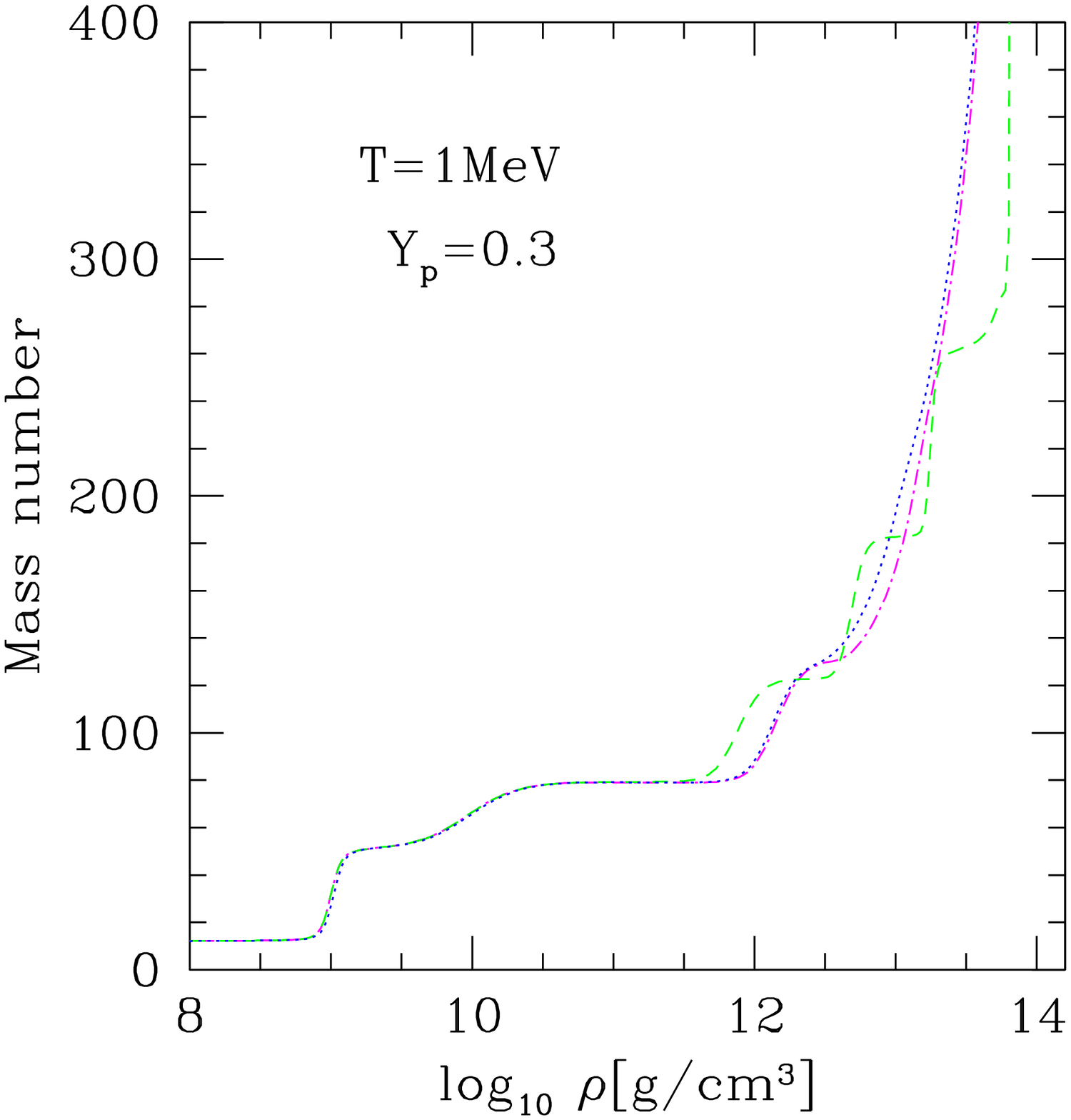}{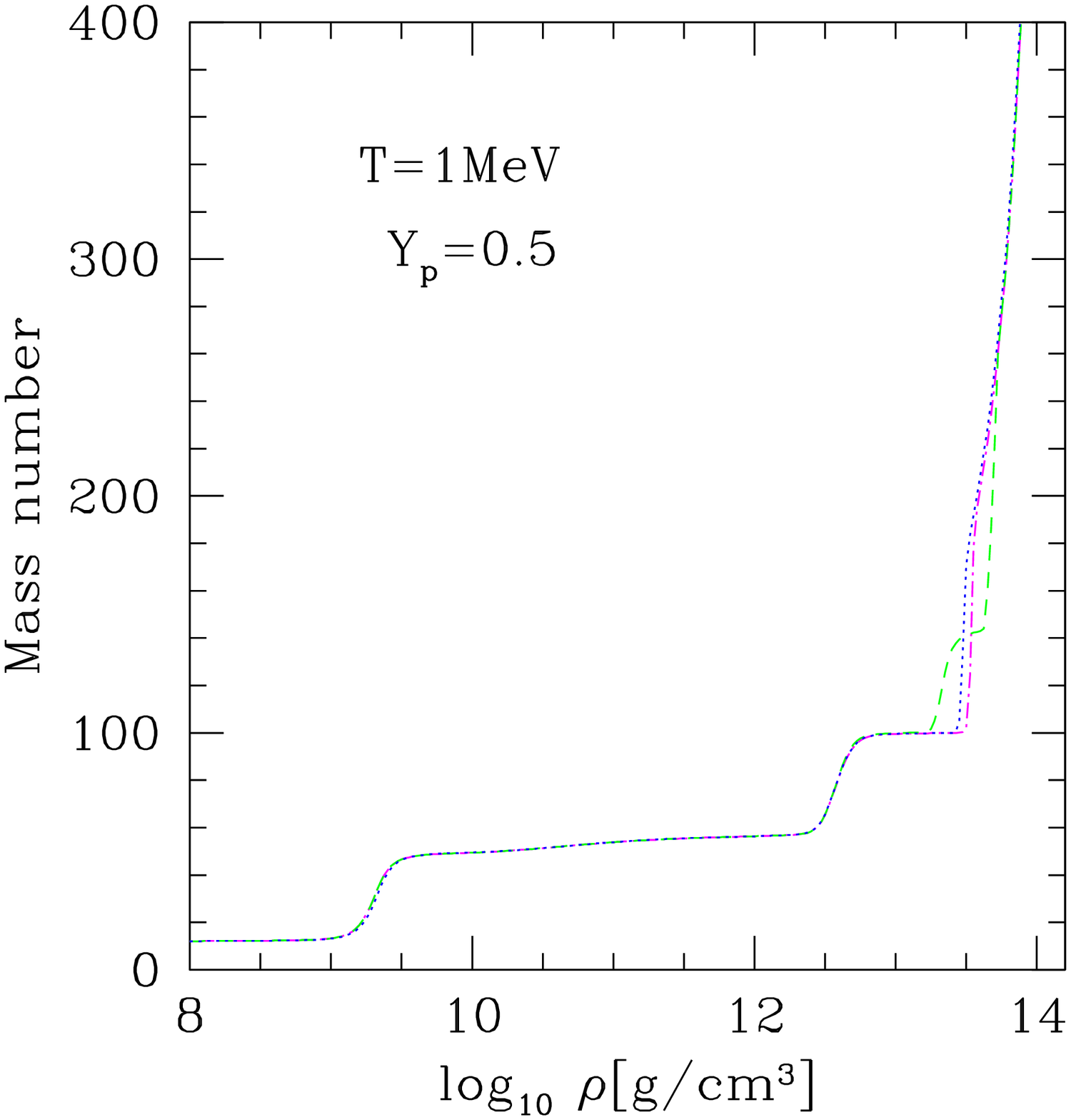} \\
   \plottwo{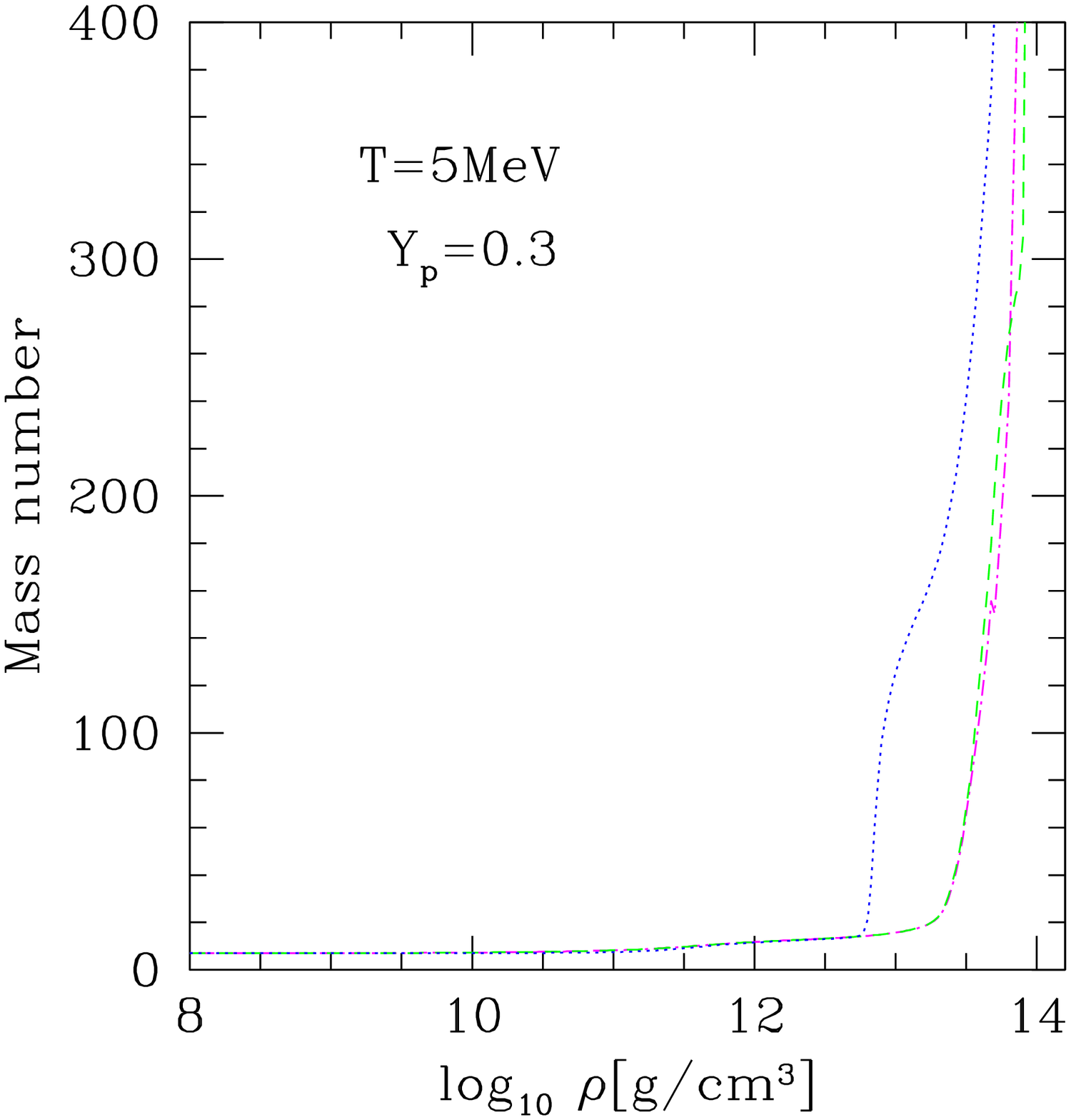}{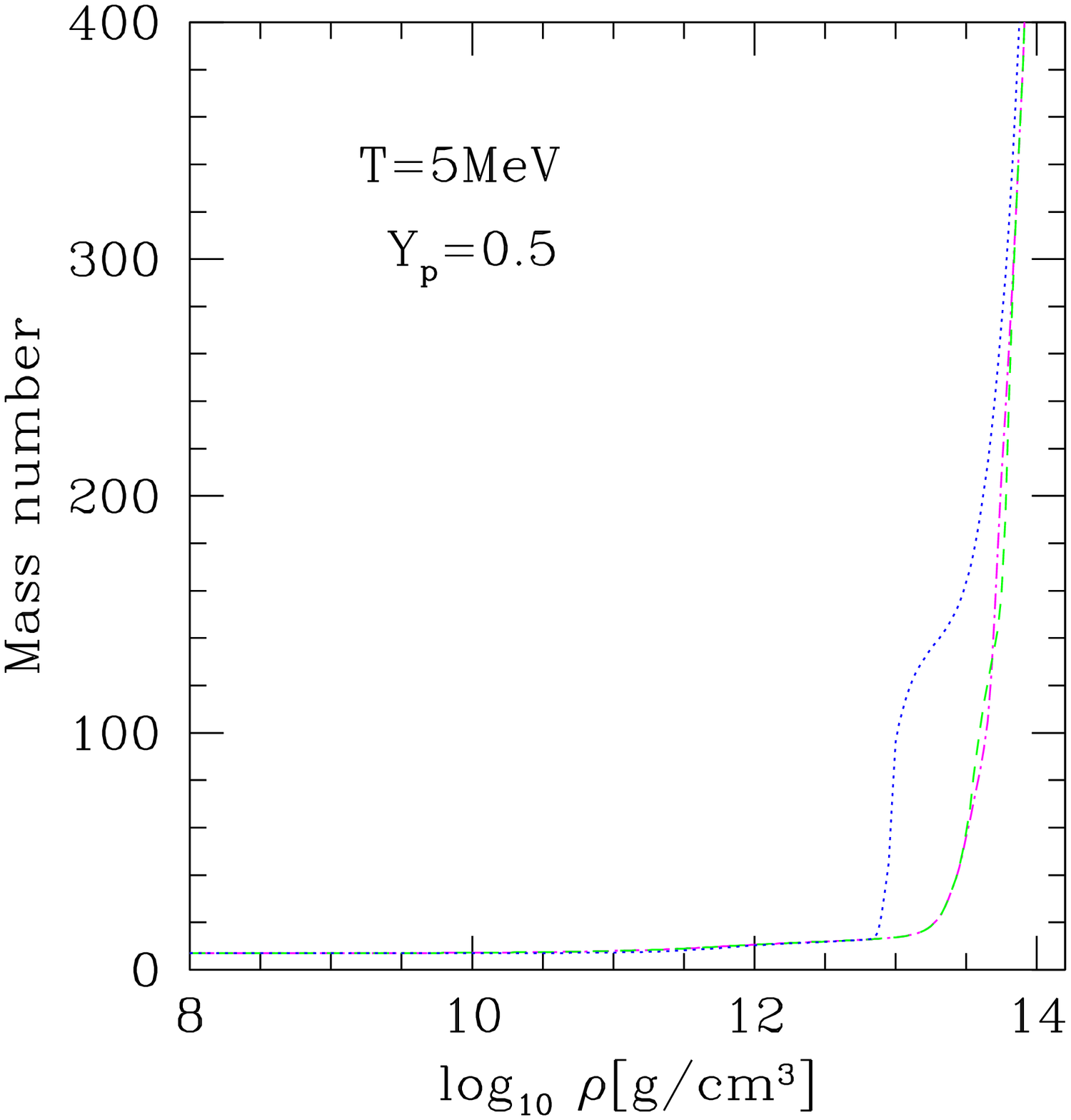} \\
   \plottwo{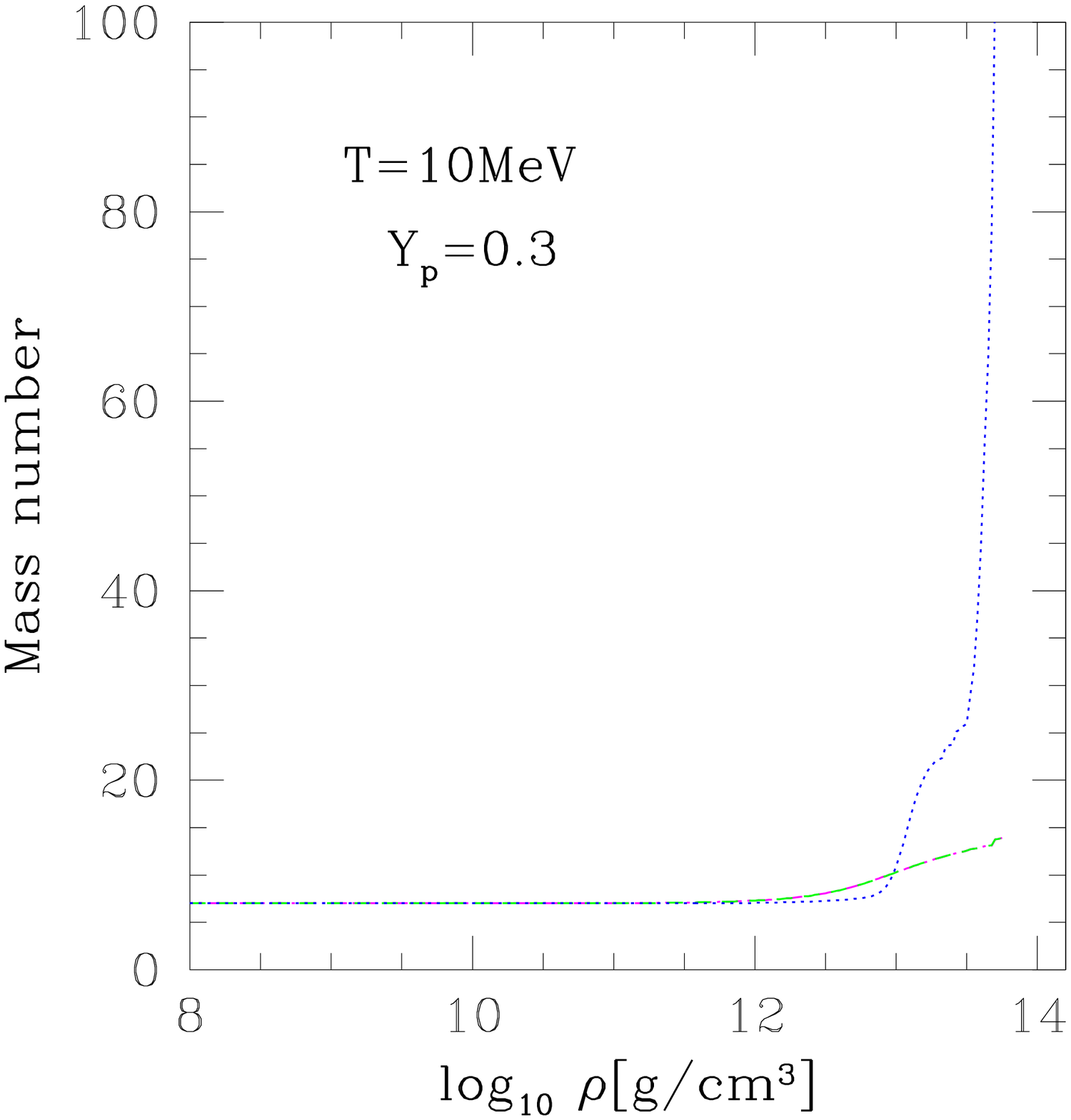}{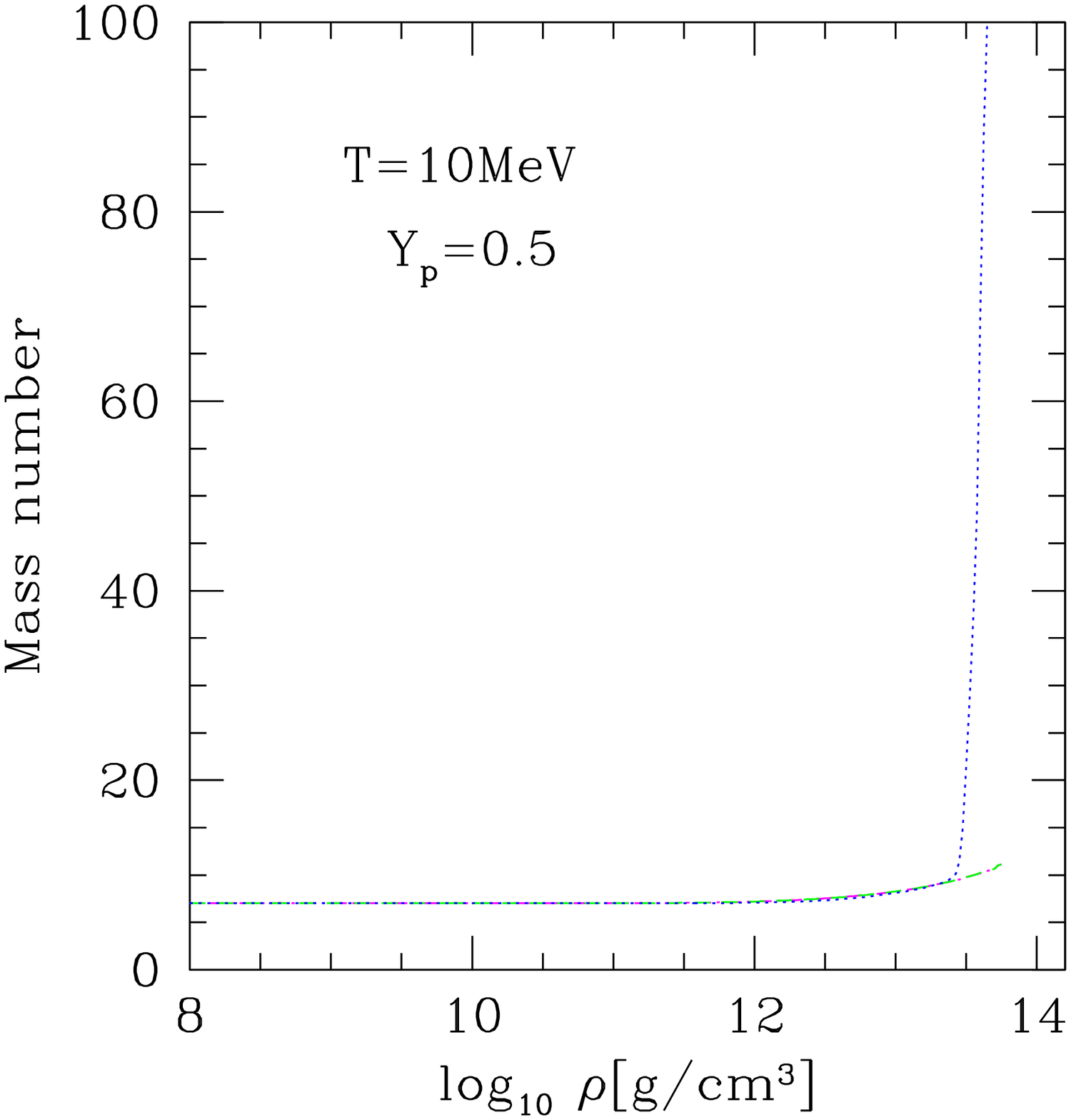} \\
    \end{tabular}
\end{center}
\caption{The average mass number of heavy nuclei with $Z>6$ for Models 0a (blue dotted lines), 1a (magenta dashed dotted lines) and 2a (green dashed lines)
as a function of density for $T=1, 5, 10 $~MeV and  $Y_p=0.3, 0.5$. }
\label{ma}
\end{figure}

We would like to stress that the new EOS provides more realistic abundances of light and heavy nuclei than the previous one.
In fact, the new EOS does not have undesirable jumps in the abundance of heavy nuclei. 
The mass estimation of light nuclei is  also more sophisticated in the new EOS.
We now briefly mention the comparison of our new EOS (Model~2b)  with others.
The detailed comparisons of our previous EOS (Model~0a) were made with EOS's employing SNA as well as with  
other multi-nuclei EOS's  in \citet{Furusawa2011}
and \citet{Buyukcizmeci}, respectively.
It is found that the EOS's with SNA give mass numbers for the representative nuclei larger than the average mass numbers
given by multi-nuclei EOS's  such as ours.
Furthermore, the two standard EOS's by \citet{Lattimer1991} and \citet{Hshen2011} lack the shell energies of nuclei
 and, as a result, show  monotonic growths of the average mass numbers of heavy nuclei,
which are in contrast with our EOS, which gives step-wise growths as shown in Fig.~\ref{ma}.
 As for light nuclei, we can provide their abundances in detail, whereas the two standard  EOS's
 with SNA give only the abundance of  alpha particles as the representative light nucleus.
We have observed that at $T=5, 10$~MeV
 the mass fractions of alpha particles given by H. Shen's EOS are larger than those in our EOS
and are smaller than the total mass fractions of all light nuclei obtained in our EOS.
This result implies that we can not neglect deuterons, tritons and helions  and the replacement of  
the ensemble of light nuclei by alpha particles is a rather poor approximation at high temperatures.

Although we can not compare our new EOS with other multi-nuclei EOS's,
it is possible to infer that our new model (Model~2b) will give
 the mass fractions of heavy nuclei similar to those obtained in Botvina's EOS \citep{Botvina04, Botvina10,Buyukcizmeci13}
 at high temperatures. 
This is because both EOS's take into account the temperature dependent bulk energies for all heavy nuclei,
as we have described in detail so far in this paper.
There should be, of course, some differences,
 which could originate from the different estimations of the surface energies and inclusions of the shell energies,
which are actually not taken into account in their EOS \citep{Buyukcizmeci}.
At low temperatures, the new EOS may give the abundances of heavy nuclei similar to those obtained by \citet{Hempel2010}.
This is because both EOS's include
 the shell effects for the neutron-rich and/or heavy nuclei by using theoretical estimations of nuclear masses.
Note, however, the difference between the theoretical mass data provided by \citet{Geng2005},
which are used in \citet{Hempel2010} and
those provided by \citet{Koura2005}, which are adopted in this paper,
may have some influences on the abundances of nuclei.
As for the light nuclei abundances, we believe that our EOS is more reliable abundances than others,
since ours takes into account the Pauli- and self-energies shifts, which could be important in medium.
It is also pointed out that in Hempel's EOS,  $g_i(T)$,
 the contribution from the internal degree of freedom to the nuclear partition function,
is also applied to light nuclei with the integration range being somewhat limited
 despite the fact that  deuterons have no excited states.
Hemepl's EOS may hence overestimate the light nuclei abundance in some cases
 although they found no significant difference from the results in more involved calculations
 by \citet{Ropke2009} and \citet{Typel2010} \citep{Hempel2011}. %rp19
We infer from \citet{Furusawa2011} and \citet{Buyukcizmeci} that thermodynamics quantities are not so different
from different EOS's  except at high densities, where the treatments of the pasta phase and abundances of light nuclei may make some differences.

There is a room for improvement in our EOS. 
The interpolation of the shell energy and the treatment of the pasta phase  are entirely
phenomenological and need justification or sophistication somehow. 
We may have to improve the EOS of uniform matter,
 which is needed to evaluate the free energy density of the free nucleons and bulk energies of heavy nuclei.
In fact, the RMF is known to have the symmetry energy larger than the canonical value,
 which will affect the neutron-richness, $Z_i /A_i$, of heavy nuclei.
In our formulation, however, it is quite simple to change the EOS of uniform nuclear matter,
once it is provided. %rp5
The combination with another EOS for supra-nuclear densities is  indeed under progress at present. 
The update of the surface tensions of nuclei, especially of neutron-rich nuclei, should be considered according
 to the progresses  in theories and experiments. %rp9
The construction of the table based on Model~2b in this paper and its application to supernova simulations are also under way. 
%%%%%%%%%%%%%%%%%%%%%%%%%%%%%%%%%%%%%%%%%%%%%%%%%%%%%%%%%%%
\acknowledgments 
We gratefully acknowledge the contribution of A. Ohnishi and C. Ishizuka to the core idea. 
We would like to thank H. Koura for useful data of the theoretical mass formula.
S.~F. would like to thank  I.N. Mishustin, G. Martinez-Pinedo, S. Typel and M. Hempel for their useful discussions.
S.~F. is grateful to FIAS for their generous support in Frankfurt.  
S.~F. is supported by the Japan Society for the Promotion of Science Research Fellowship for Young Scientists.
This work is partially supported by the Grant-in-Aid for Scientific Research on Innovative Areas (Nos. 20105004, 20105005, 24103006), 
the Grant-in-Aid for the Scientific Research (Nos. 19104006, 21540281, 22540296, 24244036) and
 the HPCI Strategic Program from the Ministry of Education, Culture, Sports, Science and Technology (MEXT) in Japan.  
A part of the numerical calculations were carried out on SR16000 at  YITP in Kyoto University.
K. S. acknowledges the usage of the supercomputers at at Research Center for Nuclear Physics (RCNP) in Osaka University, The University of Tokyo, Yukawa Institute for Theoretical Physics (YITP) in Kyoto University and High Energy Accelerator Research Organization (KEK).  
%

%%%%%%%%%%%%%%%%%%%%%%%%%%%%%%%%%%%%%%%%%%%%%%%%%%%%%%%%%%%%%%%%%%%%%%%%%%%%
%%%%%%%%%%%%%%%%%%%%%%%%%%%%%%%%%%%%%%%%%%%%%%%%%%%%%%%%%%%%%%%%%%%%%%%%%%%%
\newpage
%%%%%%%%%%%%%%%%%%%%%%
\begin{table}[t]
\begin{tabular}{ccccc}
\hline \hline
 cluster $j$ \ \ \ \ \ \ &  \ \ \ \ \ \ $a_{j,1}$ \ \ \  \ \ \ & \ \ \ \ \ \ $a_{j,2}$ \ \ \ \ \ \ & \ \ \ \ \ \ $a_{j,3}$  \ \ \ \ \ & \ \ \ \ \ \ $s_{j}$ \ \ \ \ \\
  & [MeV${}^{5/2}$fm${}^{3}$] & [MeV] & \\
 \hline
 $d$  & 38386.4 & 22.5204 & 0.2223 & 11.147\\
 $t$   & 69516.2 & 7.49232 &- & 24.575\\
 $h$  & 58442.5 & 6.07718 &- & 20.075\\
 $\alpha$ & 164371 & 10.6701 &-&49.868\\
\hline \hline
\end{tabular}
\caption{\label{tab1}%
Parameters for the quantum approach.}
\end{table}
%%%%%%%%%
%%%%%%%%%%%%%%%%%%%%%%%%%%%%%%%%%%%%%%%%%%%%%%%%%%%%%%%%%%%%%%
\begin{table}[t]
\begin{tabular}{c||c|c||c}
%\hline \hline
  \ \ \ \ \ \ model & heavy nuclei & mass data & light nuclei  \\
 \hline
 \hline
$0a$ & old LDM & Audi 2003 & old LDM  \\
 $1a$   & LDM & Audi 2003 & LDM \\
 $2a$  & LDM & Audi 2003 \& KTUY2005 & LDM \\
 $2b$ &  LDM & Audi 2003 \& KTUY2005  & quantum approach \\
$2c$ &  LDM & Audi 2003 \& KTUY2005  & mass data$+\Delta E_j^{Coul}$\\
%\hline \hline
\end{tabular}
\caption{\label{tab2}%
Different models for comparisons. 
Model~0a is the same EOS as \citet{Furusawa2011}.
The new model of heavy nuclei of Models 1a, 2a, 2b  and 2c is explained in subsection \ref{sechn}.
Audi 2003 and  KTUY2003  are the experimental and theoretical mass data respectively. 
The quantum approach of light nuclei is described in subsection \ref{secln}}
\end{table}
%%%%%%%%%%%%%%%
%%%%%%%%%%%%%%%%%%%%%%%%%%%%%%%
%%%%%%
%%%
%5
%%%%%%
%%%%%%%%%%%%%%%%%%%%%%%%%%%%555
%%%%%%%%%%%
%%%%%%%%%%%%%%%%%%%
%
%%%%%%%%
%%%%%%%%%%%%%%%%%%%%%%
%%%%%%%%%%%%%%%%%%%%%%%%%%%%%%%%%%%%%%%%%%%%%%%%%%%%%%%%%%%%%%%%%%%%%%
%%%%%%%%%%%%%%%%%%%%%%%%%%%%%%%%%%%%%%%%%%%%%%%%%%%%%%%%%%%%%%%%%%%%%%%%%%%%%%%
%
%%%%%%%%%%%%%%%%%%%%%%%%%%%%%%%%%%%%%%%%%%%%%%%%%%%%%%%%%%%%%%%%%%%%%%
%
%%%%%%%%%%%%%%%%%%%%%%%%%%%%%%%%%%%%%%%%%%%%%%%%%%%%%%%%%%%%%%%%%%%%%%
%%%
%%%%%%%%%%%%%%%%%%%%%%%%%%%%%%%%%%%%%%%%%%%%%%%%%%%%%%%%%%%%%%%%%%%%%%
%%%%%%%%%%%%%%%%%%%%%%%%%%%%%%%%%%%%%%%%%%%%%%%%%%%%%%%%%%%%%%%%%%%%%%%%%%%%%%%
%%%%%%%%%%%%%%%%%%%%%%%%%%%%%%%%%%%%%%%%%
%%%%%%%%%%%%%%%%%%%%%%%%%%%%%%%%%%
%%%%%%%%%%%%%%%%%%%
%%%%%%%%%%%%%%%%%%%%%%%%%%%%%%%%%%%%%%%%%%%%%%%%%%%%%%%%%%%%%%%%%%%
%%%%%%%%%%%%%%%%%%%%%%%%%%%%%%%%%%%%%%%%%%%%%%%%%%%%%%%%%%%%%%%%%%%%
%% \begin{figure*}
%% \vspace{15mm}
%% \epsscale{1.0}
%% \plotone{fig2.eps}
%% \caption{
%% \label{f2}}
%% \end{figure*}
\end{document}